\documentclass[journal=aaembp,manuscript=article]{achemso}

\usepackage[version=3]{mhchem} 
\usepackage{orcidlink}
\usepackage{graphicx}
\usepackage{tabularx}
\usepackage{array}
\usepackage{float}
\usepackage{ulem}
\usepackage{xcolor}
\usepackage{soul}
\usepackage[section]{placeins}
\usepackage{upgreek}
\SectionNumbersOff

\author{Md. Samrat\,\orcidlink{0009-0002-2736-9441}}
\affiliation[BUET]
{Department of Electrical and Electronic Engineering, Bangladesh University of Engineering and Technology, Dhaka, Dhaka-1205, Bangladesh}

\author{Vivek Chowdhury\,\orcidlink{0009-0006-6180-8398}}
\affiliation[BUET]
{Department of Electrical and Electronic Engineering, Bangladesh University of Engineering and Technology, Dhaka, Dhaka-1205, Bangladesh}

\author{Sake Wang\,\orcidlink{0000-0002-1004-4109}}
\affiliation[JIT]
{Department of Physics, College of Science, Jinling Institute of Technology, Nanjing, China}

\author{Ahmed Zubair\,\orcidlink{0000-0002-1833-2244}}
\email{ahmedzubair@eee.buet.ac.bd}
\affiliation[BUET]
{Department of Electrical and Electronic Engineering, Bangladesh University of Engineering and Technology, Dhaka, Dhaka-1205, Bangladesh}

\title
  {Robust Spin Splitting and Strain-Controlled Optical Response in Monolayer CrC$_2$N$_4$ for Valleytronic and Optoelectronic Applications}

\abbreviations{DFT,SOC,PBE,HSE06,RPA,PDOS}

\begin{document}

\begin{tocentry}

\centering
\includegraphics[width=8.3cm,height=4.5cm]{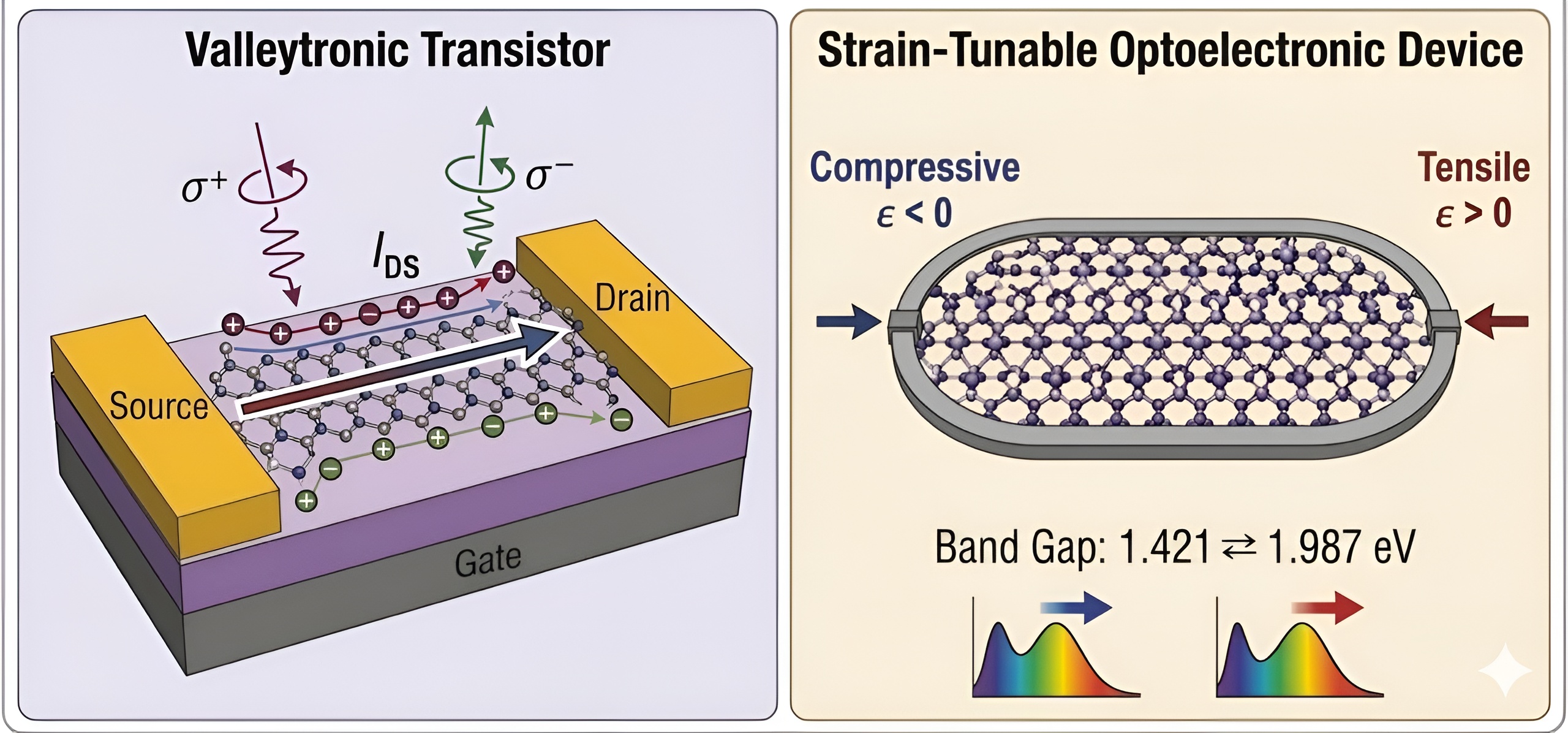}

\end{tocentry}

\begin{abstract}
Monolayer CrC$_2$N$_4$ recently emerged as a promising two-dimensional semiconductor, yet its spin--orbit-coupled (SOC) physics and strain-tunable optical response remained largely unexplored. Here, we investigated the electronic, valley, charge-transfer, and optical properties of pristine and biaxially strained monolayer CrC$_2$N$_4$ using first-principles calculations. The monolayer exhibited a direct band gap at the $\mathrm{K}$/$\mathrm{K'}$ valleys. SOC produced valley contrasting out-of-plane spin polarization, yielding a moderate valence band spin splitting of 51.9 meV and a small conduction band spin splitting of 1.7 meV. Orbital-resolved analysis showed that the edge states were mainly governed by Cr-$d$ and N-$p$ hybridization, while Bader analysis indicated polar-covalent bonding through charge transfer toward N atoms. Biaxial strain in the range of $-4\%$ to $+4\%$ tuned the band gap from 1.987 to 1.421 eV and drove an indirect-to-direct gap transition near $-1\%$ strain. Tensile strain enhanced the Berry curvature and red-shifted the optical response toward the visible--near-infrared region. These results suggested monolayer CrC$_2$N$_4$ as a promising platform for strain-engineered valleytronic and optoelectronic device applications.
\end{abstract}

\section{Introduction}

Two-dimensional (2D) materials provide a versatile platform for next-generation electronic, optoelectronic, spintronic, and valleytronic technologies because their electronic structure, optical response, carrier transport, and interfacial properties can be strongly modified at the atomic limit \cite{ahn2020,ghosh2025, chowdhury2026, islam2026}. Compared with bulk materials, atomically thin 2D semiconductors offer enhanced mechanical flexibility, reduced dielectric screening, strong light--matter interaction, and large sensitivity to external perturbations \cite{bhimanapati2015,tan2017,luo2024}. These features are especially valuable for compact device architectures in which charge, spin, photon, and momentum-space degrees of freedom can be controlled within a single material platform.

Valleytronics exploits the valley index, which originates from energetically degenerate but momentum-inequivalent extrema in the electronic band structure \cite{schaibley2016, wang2025valley}. In hexagonal 2D semiconductors, these valleys are commonly located at the time-reversal-related $\mathrm{K}$ and $\mathrm{K'}$ points of the Brillouin zone. When inversion symmetry is broken, the $\mathrm{K}$ and $\mathrm{K'}$ valleys can host opposite Berry curvatures and orbital magnetic moments, giving rise to valley-dependent Hall transport and valley-selective optical responses \cite{xiao2012,wu2019, wang2025}. In the presence of spin--orbit coupling (SOC), spin and valley degrees of freedom can become coupled, enabling valley-contrasting spin splitting and optical selection rules. Monolayer transition-metal dichalcogenides (TMDCs), such as MoS$_2$, WS$_2$, MoSe$_2$, and WSe$_2$, have served as benchmark systems for this physics because they exhibit valley-selective circular dichroism, excitonic valley polarization, spin--valley coupling, and intrinsic valley Hall effects \cite{cao2012,mak2012,wang2018}.

Although TMDCs have established the fundamental principles of 2D valley physics, practical valleytronic and optoelectronic device design requires a broader set of materials with different orbital characters, band gaps, strain responses, effective masses, and contact behavior. Mechanical strain is particularly attractive because 2D crystals can often sustain relatively large elastic deformation, allowing their band edges, orbital hybridization, SOC-induced splitting, Berry curvature, and optical transition energies to be tuned without changing chemical composition \cite{du2020,peng2020,yang2024,zollner2019}. Moreover, strain can generate valley-dependent pseudomagnetic fields that act with opposite signs in different valleys, providing an additional mechanism for manipulating valley polarization and valley-selective carrier transport\cite{fujita2010,wang2023}. In valley-active semiconductors, strain can also modify the energetic alignment of valleys at $\mathrm{K}$, $\mathrm{K'}$, $\Gamma$, and other high-symmetry points, thereby controlling direct--indirect band-gap transitions, valley splitting, and optical absorption edges \cite{wang2020strain,zhao2020,guo2024}. This strain sensitivity provides a practical route toward flexible optoelectronics, mechanically reconfigurable photodetectors, and strain-controlled valley devices.

The recent synthesis of septuple-layer MoSi$_2$N$_4$ and WSi$_2$N$_4$ has introduced a new family of 2D semiconductors beyond conventional TMDCs \cite{hong2020,novoselov2020}. These materials consist of a transition-metal nitride core protected by outer Si--N layers, giving rise to outstanding environmental stability and distinct electronic behavior. This structural motif has been generalized to the broader MA$_2$Z$_4$ family, where M is a transition metal or related element, A is typically Si or Ge, and Z is N, P, or As \cite{wanglei2021,li2023,yin2023,jin2024}. The MA$_2$Z$_4$ family offers a chemically rich design space containing semiconducting, magnetic, topological, piezoelectric, optical, and valley-related phases, making it a promising platform for multifunctional nanoscale devices \cite{chen2021,yadav2021,tho2023,latychevskaia2024}.

Valley-dependent phenomena have already been predicted in several MA$_2$Z$_4$ and their derivatives. MoSi$_2$N$_4$-family compounds can exhibit valley-contrasting Berry curvature, SOC-driven spin splitting, valley-selective circular dichroism, excitonic optical response, and strain-sensitive electronic structures \cite{li2020,ai2021,zhou2021,yangjia2021,yao2021,liu2023,xu2023}. Related studies on Janus MoSiGeZ$_4$, WSiGeZ$_4$, VSi$_2$N$_4$, MN$_2$X$_2$ (M = Mo, W; X = F, H), and CrSi$_2$N$_4$/CrSi$_2$P$_4$ further show that symmetry breaking, magnetism, SOC, and chemical substitution can be used to manipulate valley splitting, spin--valley coupling, Berry curvature, and valley-selective optical transitions in non-TMDC 2D systems \cite{guo2021,sheoran2023,cui2021,dou2020,liu2021}. These results indicate that the valley degree of freedom is not limited to TMDC monolayers, but can also be engineered in chemically diverse septuple-layer materials.

Within this broader materials landscape, monolayer CrC$_2$N$_4$ is a particularly appealing but still underexplored semiconductor. Previous first-principles calculations predicted that CrC$_2$N$_4$ is dynamically, thermally, and mechanically stable and possesses a direct semiconducting band gap, a high elastic modulus, a large tensile strength, a high lattice thermal conductivity, favorable carrier mobility, strong visible-light absorption, and a pronounced piezoelectric response \cite{mortazavi2021}. More recently, CrX$_2$N$_4$ (X = C, Si) based field-effect-transistor calculations showed that CrC$_2$N$_4$ can form an $n$-type Ohmic contact with Ti in the vertical direction, with the absence of a tunneling barrier at the CrC$_2$N$_4$--Ti interface promoting efficient electron injection \cite{shu2023}. These findings support the potential of CrC$_2$N$_4$ for device operation; however, its SOC-induced physics, orbital origin of edge states, Berry-curvature response, charge redistribution, and strain-tunable optical behavior remain insufficiently clarified.

To address these research gaps, we performed first-principles calculations to investigate pristine and biaxially strained monolayer CrC$_2$N$_4$. We analyzed its structural and electronic properties, orbital contributions at the valence band (VB) and conduction band (CB), Bader charge transfer, work function, Berry curvature, SOC-induced spin splitting, carrier effective mass, and optical response. Particular attention was given to the relationship between the orbital nature of the frontier states, valley-dependent Berry curvature, and strain-modulated optical transitions. 

\section{Methods}

In this work, first-principles calculations were carried out using the Quantum ESPRESSO package within the framework of density functional theory (DFT) \cite{Giannozzi2009}. The exchange--correlation functional was treated within the generalized gradient approximation (GGA) in the Perdew--Burke--Ernzerhof (PBE) form. Optimized norm-conserving Vanderbilt pseudopotentials were employed \cite{vanSetten2018}. The plane-wave kinetic-energy cutoffs for the wave functions and charge density were set to 110 and 440 Ry, respectively. Brillouin-zone integrations were performed using a $(12 \times 12 \times 1)$ Monkhorst--Pack \textit{k}-point mesh. The convergence of the plane-wave cutoff energy and $k$-point sampling was provided in section S1 of the supporting information. A vacuum layer of about 28 \AA\ was introduced along the out-of-plane direction to avoid spurious interactions between periodic images. Structural relaxations were performed using spin-polarized scalar-relativistic calculations and the Broyden--Fletcher--Goldfarb--Shanno (BFGS) algorithm, with the total-energy and force convergence thresholds set to $10^{-8}$ Ry and $10^{-5}$ Ry/Bohr, respectively. During relaxation, the atomic positions and in-plane lattice parameters were optimized, while the out-of-plane lattice parameter was fixed.  To obtain a more accurate electronic band structure, hybrid-functional calculations were also performed using the screened Heyd--Scuseria--Ernzerhof (HSE06) functional on the optimized structure. For the HSE06 calculations, the fraction of exact exchange was set to 0.25, and the screening parameter was taken as 0.106 bohr$^{-1}$. The exact-exchange contribution was evaluated using a $(8 \times 8 \times 1)$ $q$-point grid, and the Gygi--Baldereschi scheme was used to treat the exchange singularity. Charge transfer was analyzed using the Bader scheme \cite{Henkelman2006}.

Biaxial strain was applied in the xy-plane according to
\[
\varepsilon=\frac{a-a_0}{a_0}\times 100\%,
\]
where $a_0$ and $a$ are the equilibrium and strained lattice constants, respectively. For each strain value, the lattice constant was adjusted, and the internal atomic positions were reoptimized. Fully relativistic noncollinear calculations including spin--orbit coupling were then performed on the optimized structures. A $(12 \times 12 \times 1)$ \textit{k}-point mesh was used for the self-consistent calculations, while a $(24 \times 24 \times 1)$ mesh was used for the non-self-consistent calculations employed for Wannier interpolation. Maximally localized Wannier functions were constructed using Wannier90 \cite{Mostofi2014}, and Berry-curvature-related quantities were evaluated using WannierBerri \cite{Tsirkin2021}. The optical properties were analyzed from the frequency-dependent complex dielectric function calculated using the Yambo code within the linear-response framework \cite{sangalli2019many,marini2009yambo}. The dielectric response was obtained from the density response using the Hartree kernel, corresponding to an RPA-level optical response based on the converged DFT electronic structure. To examine the in-plane optical response, the perturbing electric field was applied along the in-plane $x$ direction. Since electron--hole interactions were not explicitly included through the Bethe--Salpeter equation, the calculated spectra were used primarily to analyze strain-dependent trends at the RPA level.

\section{Results and discussion}

\subsection{Structural Properties of Monolayer CrC$_2$N$_4$}

\begin{figure}[H]
\centering
\includegraphics[width=\textwidth]{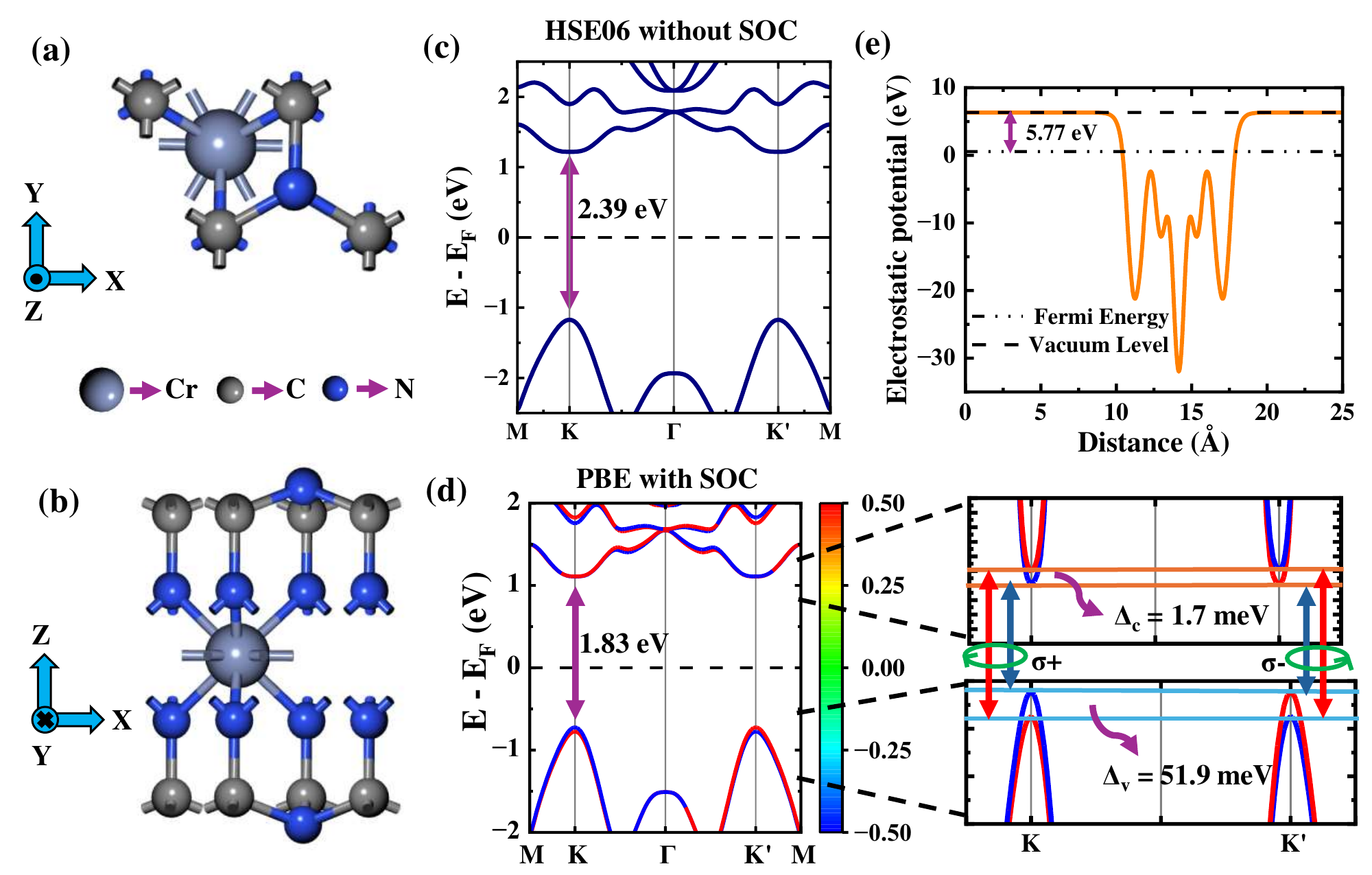}
  \caption{(a) Top view of monolayer CrC$_2$N$_4$. (b) Side view of monolayer CrC$_2$N$_4$. (c) Band structure (HSE06 without SOC) of monolayer CrC$_2$N$_4$. The band gap is marked using a purple arrow. (d) Band structure (PBE with SOC) of monolayer CrC$_2$N$_4$. The band gap is marked using a purple arrow, and CB and VB spin splittings are marked using orange and cyan horizontal lines, respectively, in the zoomed version. Red and blue arrows indicate the optical transition frequencies occurring at $\mathrm{K}$ and $\mathrm{K'}$ valleys, respectively, which are coupled via right and left-hand circular polarization of light. The color bar represents expected values of the spin operator on the spinor wave-functions varying from -0.50 (blue) to +0.50 (red) along the $z$-axis. (e) Electrostatic potential plot with respect to distance along $z$-axis. The work function of monolayer CrC$_2$N$_4$ is shown using a purple arrow.}
  \label{fig1}
\end{figure}

The optimized geometry of monolayer CrC$_2$N$_4$ is shown in \autoref{fig1}(a,b), including the top and side views. The relaxed primitive cell had a hexagonal structure consisting of seven layers (septuple layer) and can be viewed as an N-C-N-Cr-N-C-N stacking sequence along the out-of-plane direction, where the central Cr layer is sandwiched between two chemically equivalent -N-C-N sublayers. The optimized in-plane lattice constant was found to be \(a=b=2.513\) \AA, which is in excellent agreement with the previously reported value \cite{mortazavi2021}. The optimized fractional atomic coordinates were provided in section S4 of the supporting information. The previous study also reported that the structure was mechanically and thermodynamically stable. The equilibrium structural properties were listed in \autoref{tab:structural_basic}.

\begin{table}[htbp]
\centering
\caption{The structural and basic electronic parameters of monolayer CrC$_2$N$_4$. Here, $a$ is the in-plane lattice constant, $d_{\mathrm{Cr-N}}$ is the Cr--N bond length, $d_{\mathrm{C-N(in)}}$ and $d_{\mathrm{C-N(out)}}$ are the two inequivalent C--N bond lengths, $t_{\mathrm{eff}}$ is the effective thickness, and $\mathrm{E}_{\mathrm{g}}^{\mathrm{PBE}}$ and $\mathrm{E}_{\mathrm{g}}^{\mathrm{HSE06}}$ are the band gaps obtained from PBE and HSE06 calculations, respectively.}
\label{tab:structural_basic}
\footnotesize
\renewcommand{\arraystretch}{1.1}
\begin{tabular*}{\textwidth}{@{\extracolsep{\fill}}lccccccc@{}}
\hline
System & $a$ & $d_{\mathrm{Cr-N}}$ & $d_{\mathrm{C-N(in)}}$ & $d_{\mathrm{C-N(out)}}$ & $t_{\mathrm{eff}}$ & $\mathrm{E}_{\mathrm{g}}^{\mathrm{PBE}}$ & $\mathrm{E}_{\mathrm{g}}^{\mathrm{HSE06}}$ \\
 & (\AA) & (\AA) & (\AA) & (\AA) & (\AA) & (eV) & (eV) \\
\hline
Present work & 2.51 & 1.89 & 1.44 & 1.54 & 9.40 & 1.83 & 2.39 \\
Ref.~\cite{mortazavi2021} & 2.51 & 1.84 & 1.42 & 1.57 & 9.42 & 1.78 & 2.32 \\
\hline
\end{tabular*}
\end{table}

\subsection{Electronic Properties of Monolayer CrC$_2$N$_4$}

\begin{figure}[H]
\centering
\includegraphics[width=\textwidth]{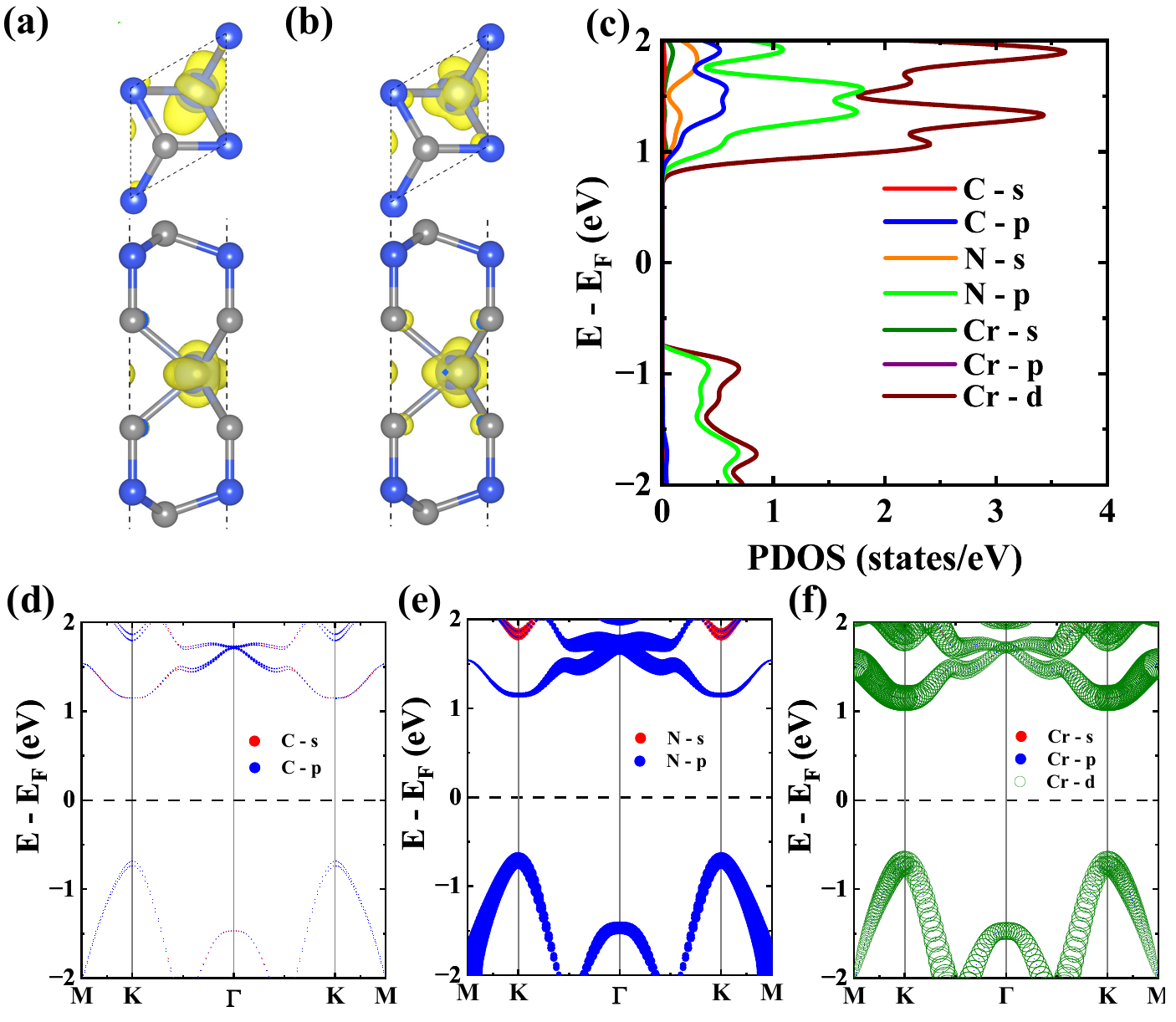}
\caption{Band-decomposed charge densities of the (a) valence band maximum (VBM) and (b) conduction band minimum (CBM) of monolayer CrC$_2$N$_4$ at the $\mathrm{K}$ point. The isosurface value is set to 0.02 \(e\,\mathrm{bohr}^{-3}\). (c) Spin--orbit-coupled projected density of states (PDOS) of monolayer CrC$_2$N$_4$, showing the orbital contributions from C, N, and Cr atoms. Orbital-projected band structures of the constituent atoms C, N, and Cr are presented in (d), (e), and (f), respectively.}
\label{fig2}
\end{figure}

The electronic structure of monolayer CrC$_2$N$_4$ (\autoref{fig1}(a,b)) was investigated through the spin--orbit-coupled (SOC) band structure, projected density of states (PDOS), and orbital-projected band dispersions, as shown in \autoref{fig1}(c,d) and \autoref{fig2}. The HSE06 band structure calculated without SOC, shown in \autoref{fig1}(c), confirmed that CrC$_2$N$_4$ was a direct band-gap semiconductor with a band gap of 2.39~eV. Both the valence band maximum (VBM) and conduction band minimum (CBM) were located at the symmetry-related points $\mathrm{K}$ and $\mathrm{K'}$ valleys. The SOC-included band structure, shown in \autoref{fig1}(d), revealed that the band gap was 1.83~eV for the PBE functional. This value was smaller than the HSE06 band gap because the PBE functional generally underestimates semiconductor band gaps. Nevertheless, the PBE with SOC calculation consistently captured the essential band characteristics and the underlying physical trends. Therefore, to reduce the computational cost while retaining the key electronic features, the PBE functional was adopted for the SOC-related calculations as a practical trade-off between accuracy and computational efficiency.

The spin-polarized band structure and PDOS without SOC were shown in section S2 of the supporting information. The spin-up and spin-down channels fully overlapped in the absence of SOC, confirming the nonmagnetic ground state of monolayer CrC$_2$N$_4$, which was further supported by the spin-resolved PDOS. After SOC was included, the semiconducting character of the system was preserved, while the spin degeneracy of the valley-edge states was lifted. As shown in \autoref{fig1}(d), the spin expectation values revealed that the band-edge states near the $\mathrm{K}$ and $\mathrm{K'}$ valleys were dominated by the out-of-plane spin component, $S_z$, whereas the in-plane components were negligible. The VBM and CBM at the $\mathrm{K}$ valley were predominantly spin-down states with $\langle S_z \rangle \approx -0.5$, whereas the corresponding states at the $\mathrm{K'}$ valley were predominantly spin-up states with $\langle S_z \rangle \approx +0.5$. The valley-dependent optical transition channels at the $\mathrm{K}$ and $\mathrm{K'}$ valleys are schematically indicated. Because the system is nonmagnetic, time-reversal symmetry is preserved. The optimized CrC$_2$N$_4$ monolayer is non-centrosymmetric, although the upper and lower sublayers are approximately related by a horizontal mirror operation; the structure lacks a spatial inversion center because inversion does not map each atomic site onto an equivalent site of the same element. The reason behind negligible valley splitting (0.2 meV at $\mathrm{K}$ and $\mathrm{K'}$) is that the transition metal Cr is not as heavy as W and Mo. In contrast, significant SOC-induced spin splittings of 51.9 meV and 1.7 meV were found at VB and CB (denoted by $\Delta_{\mathrm{v}}$ and $\Delta_{\mathrm{c}}$), respectively.  The surface electronic behavior of monolayer CrC$_2$N$_4$ is further examined through the planar-averaged electrostatic potential shown in \autoref{fig1}(e). From the difference between the vacuum level and the Fermi level, the work function is obtained as 5.77~eV. This relatively high work function suggests a strong binding of electrons to the surface and indicates that monolayer CrC$_2$N$_4$ may form stable interfaces with suitable contact materials. The work function is also relevant for carrier injection and band alignment in possible valleytronic and optoelectronic device architectures. This valley-contrasting out-of-plane spin polarization was characteristic of a Zeeman-type spin texture in a nonmagnetic system. The opposite spin polarization at the time-reversal-related $\mathrm{K}$ and $\mathrm{K'}$ valleys indicates spin--valley locking, which provides the electronic basis for the possible valley-dependent optical response and Berry curvature-driven valley Hall behavior discussed in the following sections. 

To further identify the orbital origin of the electronic states, the orbital-resolved projected density of states (PDOS) was calculated, as shown in \autoref{fig2}(c). A clear band gap was observed around the Fermi level, further confirming the semiconducting nature of monolayer CrC$_2$N$_4$. The PDOS showed that the electronic states near both the valence- and conduction band edges were mainly dominated by Cr-$d$ and N-$p$ orbitals, whereas the contribution from C orbitals was relatively weak in the vicinity of the band gap. In particular, the valence band edge was largely composed of Cr-$d$ states with significant N-$p$ hybridization, while the conduction band edge was mainly derived from Cr-$d$ states accompanied by additional N-$p$ contributions. These results indicated that the low-energy electronic behavior was governed predominantly by Cr--N orbital hybridization.

More detailed orbital information is obtained from the orbital-projected band structures shown in \autoref{fig2}(d -- f). The C-projected bands exhibit very weak intensity near the band edges, confirming the minor role of carbon in determining the valley-edge states. In contrast, the N-projected bands show appreciable contributions near both the top of the valence band and the bottom of the conduction band, demonstrating the active participation of N-$p$ orbitals in the low-energy states. The strongest contribution, however, originates from the Cr-projected bands. Quantitative orbital decomposition of the valley-edge states was provided in section S3 of the supporting information, which reveals that the VBM near the $\mathrm{K}$/$\mathrm{K'}$ valleys is mainly composed of Cr-$d_{xy}+d_{x^2-y^2}$ orbitals, with substantial hybridization from N-$p_x+p_y$ and N-$p_z$ states. In contrast, the CBM is predominantly governed by Cr-$d_{z^2}$ and Cr-$d_{xz}+d_{yz}$ orbitals, accompanied by a noticeable N-$p_z$ contribution. This pronounced Cr-$d$/N-$p$ hybridization provides the microscopic origin of the valley-edge electronic structure and plays a key role in the spin--valley-coupled properties of monolayer CrC$_2$N$_4$. The spin character and orbital contributions of the valley-edge states are summarized in \autoref{tab:valley_spin_orbital}.

\begin{table}[htbp]
\centering
\caption{Valley-resolved spin character and dominant orbital contributions (\%) of the band-edge states of monolayer CrC$_2$N$_4$.}
\label{tab:valley_spin_orbital}
\footnotesize
\renewcommand{\arraystretch}{1.15}
\begin{tabular*}{\textwidth}{@{\extracolsep{\fill}}lcccccccc@{}}
\hline
State & Valley & Spin & Cr-$d_{xy}+d_{x^2-y^2}$ & Cr-$d_{z^2}$ & Cr-$d_{xz}+d_{yz}$ & N-$p_z$ & N-$p_x+p_y$ \\ 
\hline
VBM & $\mathrm{K}$ & Down &  63.03\% & -- & -- & 7.31\% & 29.42\% \\ 
CBM & $\mathrm{K}$ & Down &  -- & 41.99\% & 39.24\% & 6.76\% & 9.21\% \\ 
VBM & $\mathrm{K'}$ & Up &  62.99\% & -- & -- & 7.32\% & 29.45\% \\ 
CBM & $\mathrm{K'}$ & Up &  -- & 41.99\% & 39.23\% & 6.76\% & 9.21\% \\ 
\hline
\end{tabular*}
\end{table}

\subsection{Charge density and Bader charge analysis}

\autoref{fig2} (a, b) show the band-decomposed charge densities of the valence band maximum (VBM) and conduction band minimum (CBM) of monolayer CrC$_2$N$_4$ at the $\mathrm{K}$ point. The band-edge charge densities were mainly localized around the Cr atom and its neighboring N atoms, whereas the contribution from C atoms was comparatively weak. This was consistent with the orbital-resolved electronic structure, where the frontier states are found to be dominated primarily by Cr-$d$ and N-$p$ orbitals.

To further examine the bonding nature, Bader charge analysis was performed using the converged total charge density. The results indicated a clear charge transfer from Cr and C atoms to N atoms. In particular, the Cr atom lost \(1.374\,e\), while the two C atoms together lost \(2.345\,e\). In contrast, the two outer N atoms gained \(1.480\,e\), and the two inner N atoms gained \(2.239\,e\). The larger charge accumulation on the inner N atoms suggested their stronger participation in the Cr--N bonding environment. These results indicated that bonding in monolayer CrC$_2$N$_4$ had a significant polar-covalent character, arising from electron transfer from Cr and C atoms toward the more electronegative N atoms.

\subsection{\textcolor{black}{Valley Physics: Berry Curvature and Strain Effects}}

The Berry curvature is a fundamental quantity that describes the geometrical property of electronic Bloch states in momentum space and acts as an effective magnetic field in momentum space when time-reversal symmetry or spatial inversion symmetry is broken~\cite{wang2025valley,wang2020strain}. Following the valley effective-model description of Li \textit{et al.}~\cite{li2020}, the low-energy electronic states near the inequivalent valleys can be described by a two-band Dirac-type Hamiltonian. In the absence of spin--orbit coupling (SOC), the Hamiltonian around the valley index $\tau=\pm1$, corresponding to the $\mathrm{K}$ and $\mathrm{K'}$ valleys, is written as
\begin{equation}
H_0^\tau
=
\alpha \left(\tau k_x \sigma_x + k_y \sigma_y\right)
+
\frac{\Delta}{2}\sigma_z ,
\label{eq:effective_hamiltonian_no_soc}
\end{equation}
where $\mathbf{k}$ is measured from the valley center, $\sigma_i$ are the Pauli matrices in the two-orbital basis, $\alpha$ is the effective coupling parameter, and $\Delta$ is the band-gap parameter. When SOC is included, the effective Hamiltonian becomes
\begin{equation}
H^\tau
=
\alpha \left(\tau k_x \sigma_x + k_y \sigma_y\right)
+
\frac{\Delta}{2}\sigma_z
-
\frac{\lambda}{2}\tau\left(\sigma_z-1\right)\hat{s}_z ,
\label{eq:effective_hamiltonian_soc}
\end{equation}
where $\lambda$ denotes the effective SOC strength and $\hat{s}_z$ is the Pauli matrix for spin. This form captures the essential spin--valley-coupled character of the band-edge states, with stronger SOC-induced splitting in the valence band and much weaker splitting in the conduction band.

For a two-dimensional system, the Berry curvature has only the out-of-plane component and behaves as a pseudoscalar. For the $n$th band at wave vector $\mathbf{k}$, it can be written in the Kubo-like form as~\cite{li2020}
\begin{equation}
\Omega_{n,z}(\mathbf{k})
=
-2\,\mathrm{Im}
\sum_{m\neq n}
\frac{
\left\langle n\mathbf{k} \middle| \hat{v}_x \middle| m\mathbf{k} \right\rangle
\left\langle m\mathbf{k} \middle| \hat{v}_y \middle| n\mathbf{k} \right\rangle
}{
\left(\omega_m-\omega_n\right)^2
},
\label{eq:berry_curvature_kubo}
\end{equation}
where $\hat{v}_x$ and $\hat{v}_y$ are the velocity operators, and $E_n=\hbar\omega_n$ is the energy of the state $\left|n\mathbf{k}\right\rangle$. The velocity operators can be obtained from the effective Hamiltonian as
\begin{equation}
\hat{v}_i
=
\frac{1}{\hbar}
\frac{\partial H^\tau}{\partial k_i},
\qquad i=x,y .
\label{eq:velocity_operator}
\end{equation}

Using the SOC-included effective Hamiltonian in~\autoref{eq:effective_hamiltonian_soc}, the Berry curvature for the conduction band at valley $\tau$ and spin eigenvalue $s=\pm1$ is given by~\cite{li2020}
\begin{equation}
\Omega_c^\tau(\mathbf{k},s)
=
-\tau
\frac{
2\alpha^2\widetilde{\Delta}_{\tau s}
}{
\left(
\widetilde{\Delta}_{\tau s}^{\,2}
+
4\alpha^2 k^2
\right)^{3/2}
},
\label{eq:berry_curvature_conduction}
\end{equation}
where
\begin{equation}
\widetilde{\Delta}_{\tau s}
=
\Delta-\tau s\lambda .
\label{eq:delta_tilde}
\end{equation}
For the corresponding spin-split valence band, the Berry curvature has the opposite sign,
\begin{equation}
\Omega_v^\tau(\mathbf{k},s)
=
-\Omega_c^\tau(\mathbf{k},s).
\label{eq:berry_curvature_valence}
\end{equation}
~\autoref{eq:berry_curvature_conduction} shows that the Berry curvature changes sign between the two time-reversal-related valleys because of the factor $\tau$. Therefore, the $\mathrm{K}$ and $\mathrm{K'}$ valleys possess nonzero Berry curvatures with opposite signs, and the magnitude is maximum at the valley center, $k=0$.

In the presence of an external in-plane electric field, the nonzero Berry curvature gives rise to an anomalous transverse velocity, which is responsible for Hall-type responses such as the valley Hall effect. Since monolayer CrC$_2$N$_4$ preserves time-reversal symmetry, the total charge Hall response is expected to cancel when both valleys are equally populated. However, the opposite Berry curvatures at the $\mathrm{K}$ and $\mathrm{K'}$ valleys can drive carriers from the two valleys toward opposite transverse directions, producing a pure valley Hall current. This valley-contrasting behavior can be summarized as
\begin{equation}
\Omega_z(\mathrm{K})
=
-\Omega_z(\mathrm{K'}).
\label{eq:valley_contrasting_berry_curvature}
\end{equation}
Therefore, even when the net charge Hall current is zero, opposite anomalous velocities in the two valleys can generate a valley-polarized Hall response.

\begin{figure}[H]
    \centering
    \includegraphics[width=\textwidth]{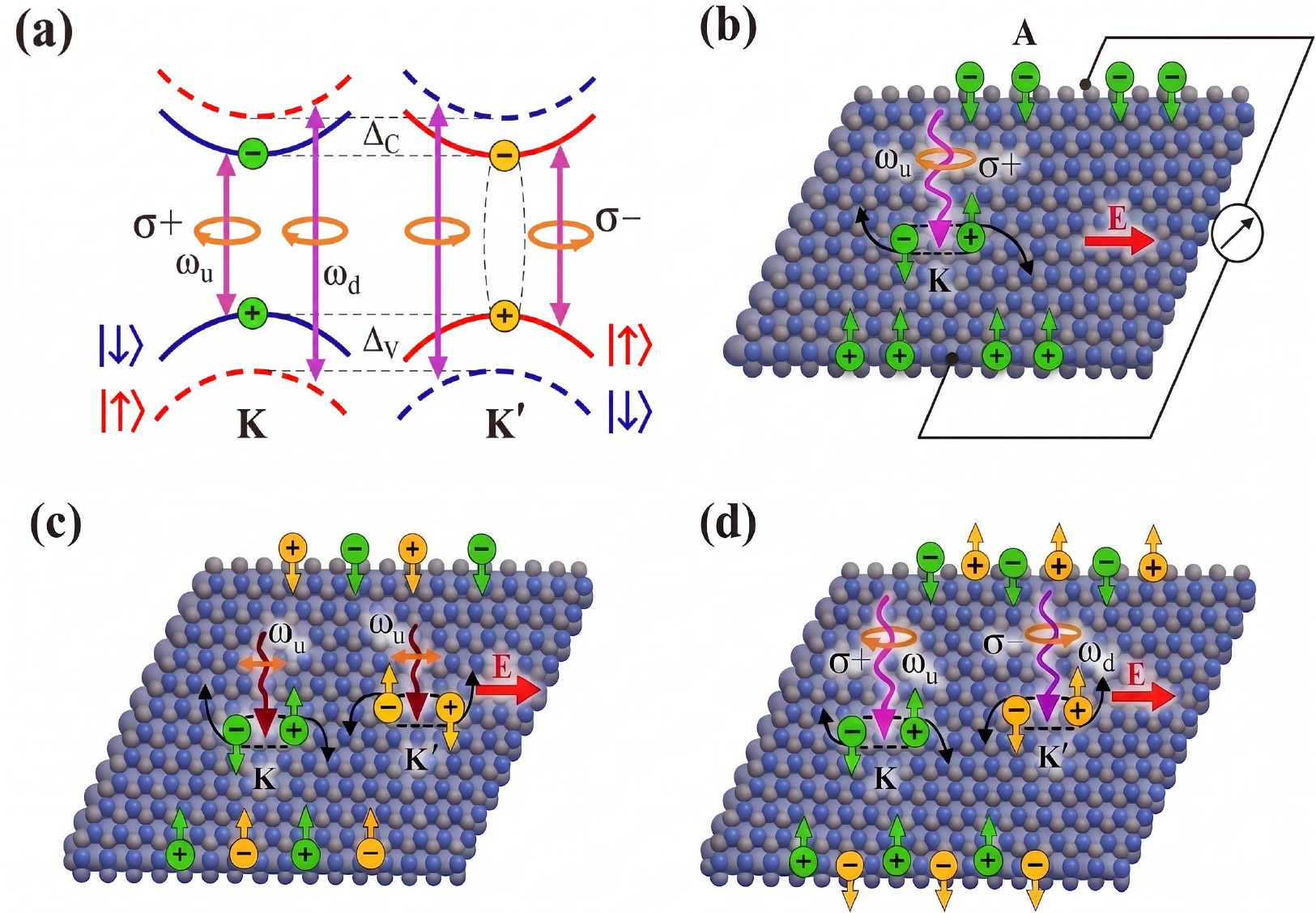}
   \caption{Spin-- and valley-coupled optical selection rules and valley-dependent Hall response. (a) Valley-resolved interband optical transitions at the $\mathrm{K}$ and $\mathrm{K}^{\prime}$ points under circularly polarized light, $\sigma^{+}$ and $\sigma^{-}$. The transition energies $\omega_\mathrm{u}$ and $\omega_\mathrm{d}$ and the SOC-induced band splittings $\Delta_\mathrm{c}$ and $\Delta_\mathrm{v}$ in the conduction and valence bands are indicated. The red and blue parabolas denote out-of-plane spin-up and spin-down states, respectively. (b) Circularly polarized light induced a photoinduced charge Hall effect by selectively exciting carriers in one valley under an applied in-plane electric field E. (c) Linearly polarized light induces spin and valley Hall effects of electrons and holes, with an absent charge Hall effect. (d) $\upsigma^{+}$ polarized light with frequency $\omega_\text u$ and $\upsigma^{-}$ polarized light with frequency $\omega_\text d$ excite spin-down electrons and spin-up holes in both valleys.}
    \label{fig:hall_effect}
\end{figure}

The spin and valley-coupled transport mechanism in \autoref{fig:hall_effect} followed from valley-selective optical excitation and opposite Berry curvatures at the $\mathrm{K}$ and $\mathrm{K'}$ valleys~\cite{xiao2012}. Because monolayer CrC$_2$N$_4$ lacked inversion symmetry and exhibited SOC-induced spin splitting, the valley-edge states were spin--valley locked: the band-edge states at $\mathrm{K}$ were mainly spin-down, whereas those at $\mathrm{K'}$ were mainly spin-up, as shown in \autoref{fig1}(d) and \autoref{tab:valley_spin_orbital}. Circularly polarized excitation selectively populated one valley and, under an in-plane electric field $\mathbf{E}$, the Berry-curvature-induced anomalous velocity $\mathbf{v}_a=-(e/\hbar)\mathbf{E}\times\boldsymbol{\Omega}(\mathbf{k})$ gave rise to a photoinduced charge Hall response. In contrast, linearly polarized excitation populated both valleys. Because time-reversal symmetry required $\Omega_z(\mathrm{K})=-\Omega_z(\mathrm{K'})$, the net charge Hall current canceled, while spin- and valley-dependent transverse carrier separation remained. A two-color excitation using $\upsigma^+(\omega_u)$ and $\upsigma^-(\omega_d)$ channels further generated spin-- and valley-polarized carriers in both valleys, providing an optical route to control valley-dependent Hall transport in monolayer CrC$_2$N$_4$.

\subsection{Effect of Biaxial Strain on Band Structure, Berry Curvature, Spin Splitting, and Optical Property}

Mechanical strain modifies the lattice geometry and changes the bond lengths between neighboring atoms. Since the hopping amplitudes in a tight-binding Hamiltonian depend sensitively on the bond lengths, strain changes the electronic band structure, the Bloch wave functions, and therefore the Berry curvature. For uniaxial strain, the strain direction is described by an angle $\alpha$ with respect to the positive $x$-axis. The strain tensor can be written as
\begin{equation}
\boldsymbol{\varepsilon}
=
\begin{pmatrix}
\varepsilon_{xx} & \varepsilon_{xy}\\
\varepsilon_{yx} & \varepsilon_{yy}
\end{pmatrix},
\end{equation}
where
\begin{align}
\varepsilon_{xx} &= \varepsilon_0\left(\cos^2\alpha-\nu\sin^2\alpha\right),\\
\varepsilon_{yy} &= \varepsilon_0\left(\sin^2\alpha-\nu\cos^2\alpha\right),\\
\varepsilon_{xy} &= \varepsilon_{yx}
= \varepsilon_0(1+\nu)\sin\alpha\cos\alpha .
\end{align}
Here, $\varepsilon_0$ is the magnitude of the applied uniaxial strain and $\nu$ is Poisson's ratio.

\begin{figure}[H]
\centering
\includegraphics[width=\textwidth]{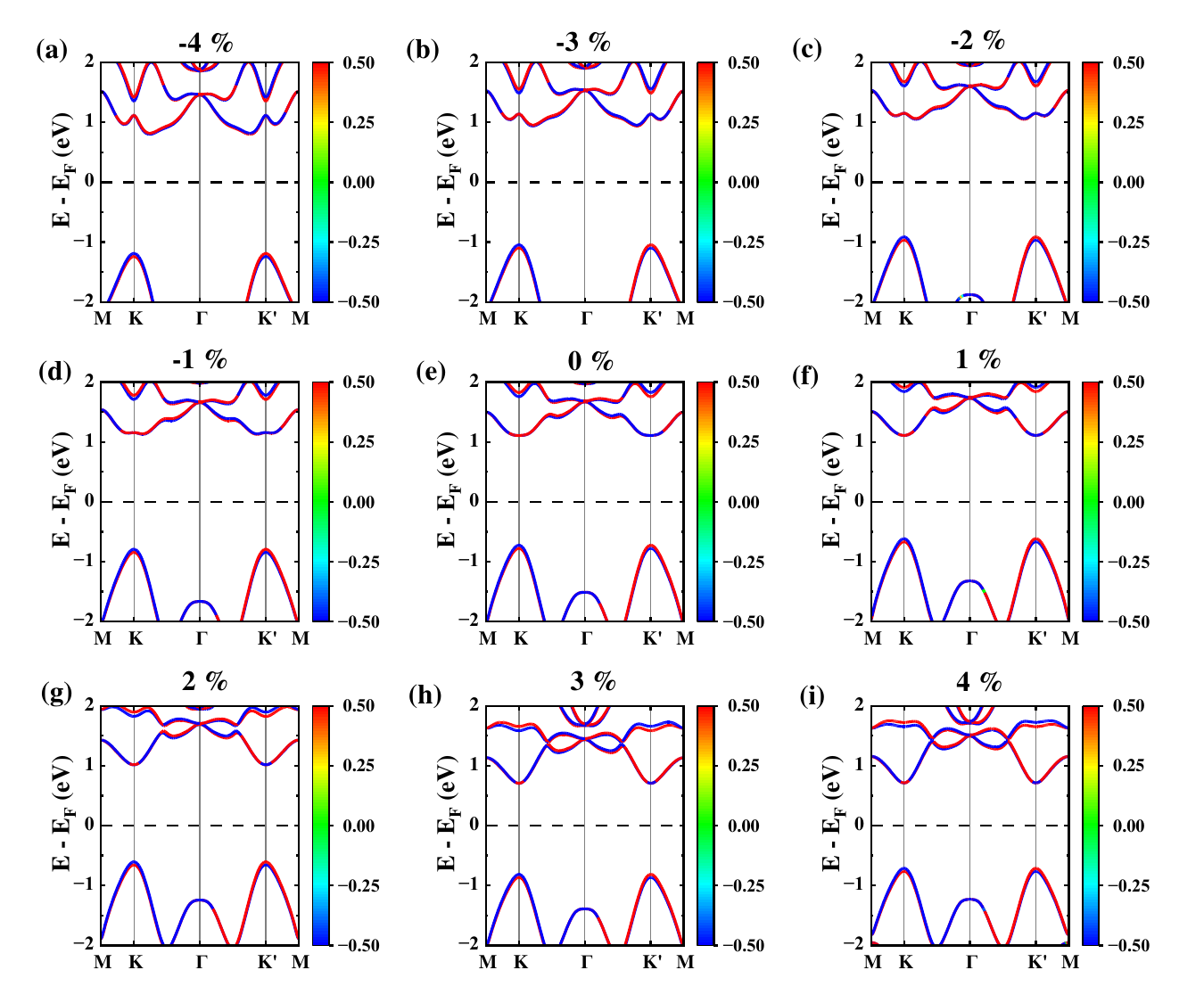}
  \caption{Band structures (with SOC) for (a) $-4\%$, (b) $-3\%$, (c) $-2\%$, (d) $-1\%$, (e) $0\%$, (f) $1\%$, (g) $2\%$, (h) $3\%$, and (i) $4\%$ biaxial strain. The color bar represents expected values of the spin operator on the spinor wave-functions varying from -0.50 (blue) to +0.50 (red) along the $z$-axis. } 
  \label{fig3}
\end{figure}

\autoref{fig3}(a--i) illustrate the variation in the band structure of monolayer CrC$_2$N$_4$ under applied biaxial strain ranging from $-4\%$ to $4\%$. It was observed that the CB was strongly modulated by biaxial strain, whereas the VB remained the same. Compressive biaxial strain induced indirect band gaps, whereas the unstrained and tensile-strained cases exhibited direct band gaps. Moreover, the spin-up and spin-down bands remained symmetric with respect to each other for all strain values, indicating the absence of net magnetization in the system.

\begin{figure}[H]
\centering
\includegraphics[width=\textwidth]{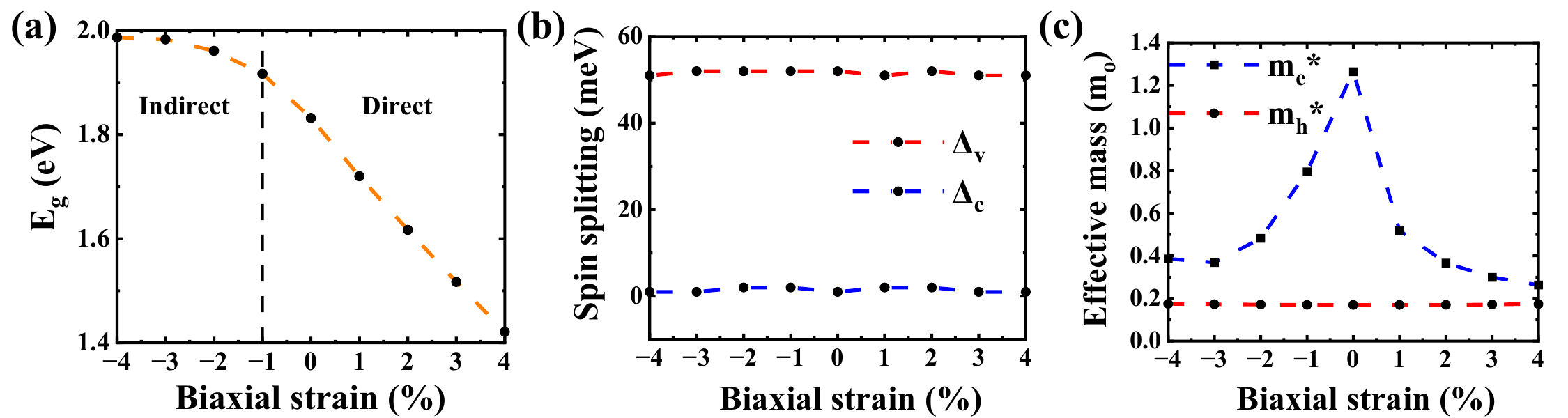}
\caption{(a) Band gap, $\mathrm{E}_{\mathrm{g}}$, (b) CB spin splitting, $\Delta_{\mathrm{c}}$ and VB spin splitting, $\Delta_{\mathrm{v}}$, and (c) effective mass variation of electron ($\mathrm{m_e \ast}$) and hole ($\mathrm{m_h \ast}$) in terms of electron rest mass ($\mathrm{m_o}$) with respect to biaxial strain.}
\label{trend}
\end{figure}

\autoref{trend}(a--c) illustrate the variations of the band gap, $\mathrm{E}_{\mathrm{g}}$, spin splitting, and carrier effective masses under biaxial strain from $-4\%$ to $4\%$. As shown in \autoref{trend}(a), the band gap decreased monotonically with increasing biaxial strain, falling from 1.987~eV at $-4\%$ to 1.421~eV at $4\%$. A transition from an indirect to a direct band gap was observed near $-1\%$ strain, indicating that tensile strain favored the direct-gap nature of the system. The spin splitting results in \autoref{trend}(b) revealed a clear contrast between the valence and conduction bands. The VB spin splitting, $\Delta_{\mathrm{v}}$, remained nearly unchanged throughout the studied strain range, with values confined to 51--52~meV, whereas the CB spin splitting, $\Delta_{\mathrm{c}}$, stayed very small, varying only between 1 and 2~meV. This behavior suggested that the spin polarization was much more pronounced in the valence band than in the conduction band and was only weakly influenced by biaxial strain. The strain dependence of the effective masses is presented in \autoref{trend}(c). The electron effective mass, $\mathrm{m}_{\mathrm{e}}^{\ast}$, exhibited a strong sensitivity to strain, increasing from $0.386\,\mathrm{m}_{0}$ at $-4\%$ to a maximum value of $1.265\,\mathrm{m}_{0}$ at 0\% strain, and then decreasing sharply to $0.263\,\mathrm{m}_{0}$ at $4\%$ strain. In contrast, the hole effective mass, $\mathrm{m}_{\mathrm{h}}^{\ast}$, remained almost constant over the entire strain range, varying only slightly between $0.169\,\mathrm{m}_{0}$ and $0.175\,\mathrm{m}_{0}$. Overall, these results demonstrated that biaxial strain strongly tuned the electronic structure of monolayer CrC$_2$N$_4$, particularly the band gap and electron effective mass, while leaving the spin splitting and hole effective mass nearly unaffected.

In semiconducting valley materials such as transition-metal dichalcogenides, biaxial tensile strain may reduce or increase the band gap depending on the orbital character of the band edges. Since Berry curvature is usually enhanced when the band gap becomes smaller, tensile biaxial strain can increase the Berry curvature magnitude near $\mathrm{K}$ and $\mathrm{K'}$ if it reduces the direct gap. Conversely, compressive biaxial strain can suppress the Berry curvature if it increases the gap. The opposite signs of Berry curvature at $\mathrm{K}$ and $\mathrm{K'}$ are expected to remain as long as time-reversal symmetry is preserved:
\begin{equation}
\Omega_z^{\mathrm{K}}(\mathbf{k})
=
-\Omega_z^{\mathrm{K'}}(-\mathbf{k}) .
\end{equation}

\begin{figure}[H]
\centering
\includegraphics[width=\textwidth]{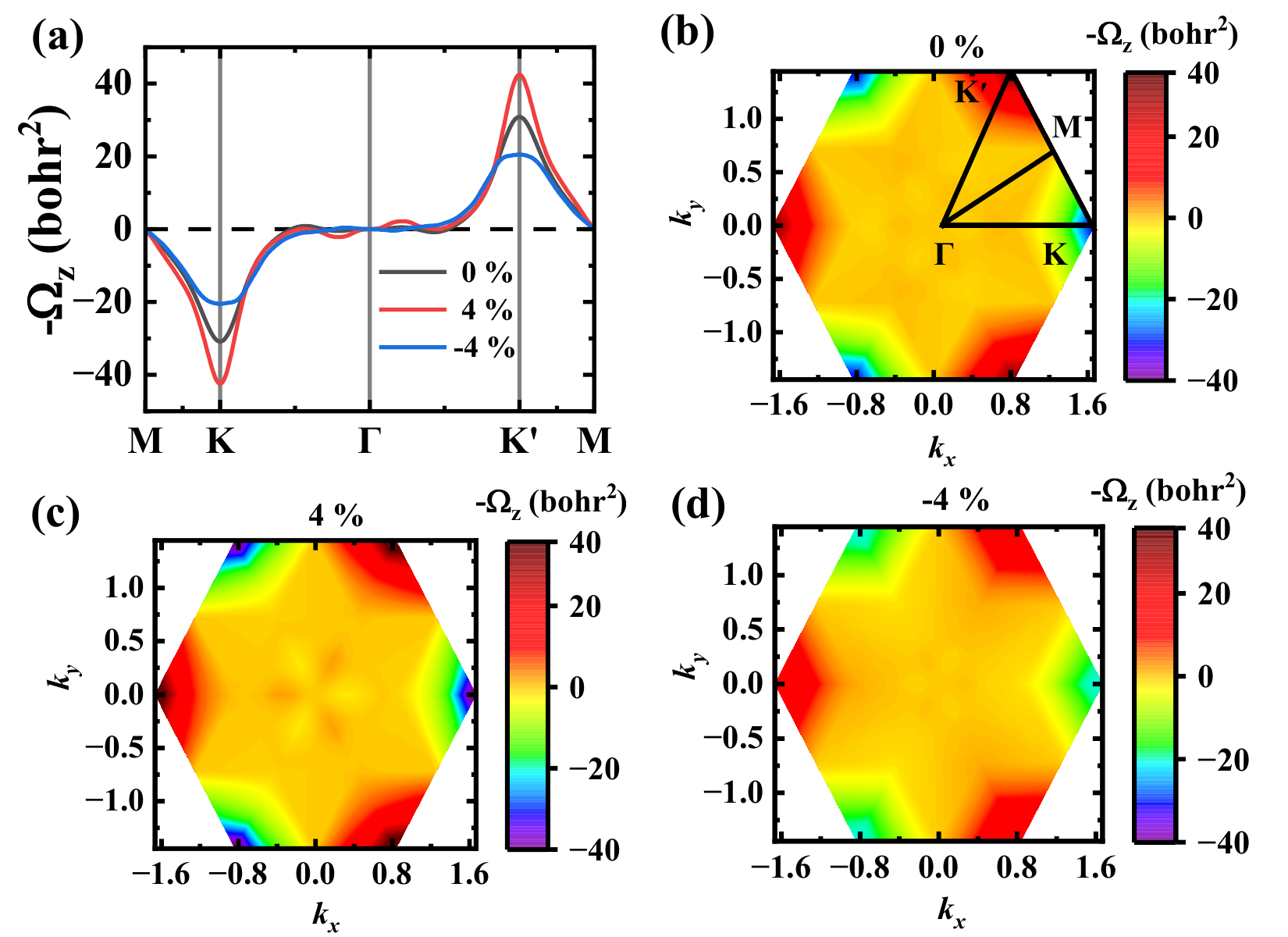}
  \caption{(a) Berry curvature, $-\mathrm{\Omega}_{\mathrm{z}}\,(\text{bohr}^2)$
    along the path $\mathrm{M} \to \mathrm{K} \to \mathrm{\Gamma} \to \mathrm{K'} \to \mathrm{M}$, showing opposite magnitudes at $\mathrm{K}$ and $\mathrm{K'}$ valleys for different biaxial strains (0 \%, 4\%, and -4\%). A contour plot for Berry curvature, $-\mathrm{\Omega}_{\mathrm{z}}\,(\text{bohr}^2)$ demonstrating contrasting behavior at alternating $\mathrm{K}$ and $\mathrm{K'}$ valleys over the whole hexagonal Brillouin zone for biaxial strains (b) 0 \%, (c) 4 \%, and (d) -4 \%. The colormap indicates the value of Berry curvature, $-\mathrm{\Omega}_{\mathrm{z}}$ ranging from -40 $\text{bohr}^2$ to 40 $\text{bohr}^2$. } 
  \label{fig5}
\end{figure}

Therefore, biaxial strain is expected mainly to tune the magnitude of Berry curvature, while uniaxial strain can additionally distort and shift the Berry curvature distribution in momentum space. Since monolayer CrC$_2$N$_4$ lacks spatial inversion symmetry while preserving time-reversal symmetry, finite local Berry curvature is allowed at $\mathrm{K}$ and $\mathrm{K'}$ valleys. \autoref{fig5}(a) depicts the Berry curvature, $-\mathrm{\Omega}_{\mathrm{z}}$  of monolayer CrC$_2$N$_4$ along the high symmetry path $\mathrm{M} \to \mathrm{K} \to \mathrm{\Gamma} \to \mathrm{K'} \to \mathrm{M}$ for three different biaxial strain levels (0 \%, 4 \%, -4 \%). Here, the tensile biaxial strain (4 \%) enhanced the Berry curvature to 42.35 $\text{bohr}^2$ from 30.85 $\text{bohr}^2$ (unstrained case), whereas the compressive biaxial strain (-4 \%) lowered the value to 20.52 $\text{bohr}^2$ at $\mathrm{K}$/$\mathrm{K'}$ valleys. \autoref{fig5}(b-d) represent the contour plots of 0\%, 4\%, and -4\% cases respectively, where the strain modulation can be seen at the vertices of the hexagons. The tensile biaxial strain also created a faint wiggle (opposite in sign with respect to $\mathrm{K}$ and $\mathrm{K'}$ valleys) near the $\mathrm{\Gamma}$ point, which was negligible for the other cases.

The optical response of a strained monolayer is governed by the complex dielectric function,
\begin{equation}
\varepsilon(\omega)=\varepsilon_1(\omega)+i\varepsilon_2(\omega),
\end{equation}
where $\varepsilon_1(\omega)$ and $\varepsilon_2(\omega)$ are the real and imaginary parts of the dielectric function, respectively. The refractive index, $\mathrm{n}(\omega)$, and extinction coefficient, $\mathrm{k}(\omega)$, are related to the dielectric function as
\begin{equation}
\mathrm{n}(\omega)=
\left[
\frac{\sqrt{\varepsilon_1^2(\omega)+\varepsilon_2^2(\omega)}
+\varepsilon_1(\omega)}{2}
\right]^{1/2},
\end{equation}
\begin{equation}
\mathrm{k}(\omega)=
\left[
\frac{\sqrt{\varepsilon_1^2(\omega)+\varepsilon_2^2(\omega)}
-\varepsilon_1(\omega)}{2}
\right]^{1/2}.
\end{equation}
The absorption coefficient, $\upalpha(\omega)$, can then be expressed as
\begin{equation}
\upalpha(\omega)=\frac{2\omega \mathrm{k}(\omega)}{c},
\end{equation}
where $\omega$ is the angular frequency of the incident photon and $c$ is the speed of light. Since the photon energy and wavelength are related by
\begin{equation}
E=\frac{hc}{\lambda},
\end{equation}
any strain-induced change in the electronic band gap and interband transition energies directly modifies the absorption edge and optical peak positions. In general, tensile strain tends to reduce the band gap and shifts optical transitions toward longer wavelengths, whereas compressive strain usually increases the transition energy and shifts the optical response toward shorter wavelengths.

\begin{figure}[H]
\centering
\includegraphics[width=\textwidth]{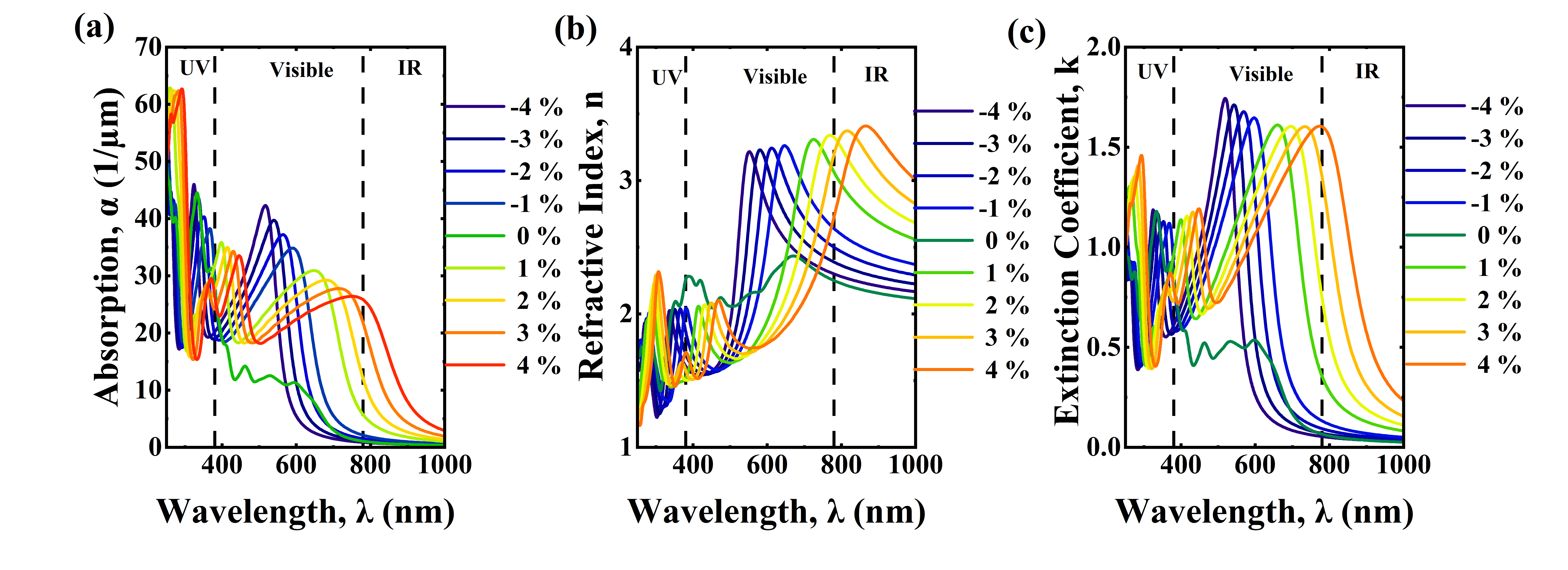}
  \caption{(a) Absorption, $\alpha$, (b) Refractive index, n, and (c) Extinction coefficient, k with respect to wavelength, $\lambda$ for $-4\%$ to $4\%$ biaxial strain.} 
  \label{fig4}
\end{figure}

\autoref{fig4}(a--c) show the variation of absorption coefficient, $\upalpha$, refractive index, $\mathrm{n}$, and extinction coefficient, $\mathrm{k}$, with wavelength, $\lambda$, under biaxial strain from $-4\%$ to $4\%$. The absorption spectra showed strong optical activity in the UV region, where the absorption coefficient reached its highest value of about $60~\mu\mathrm{m}^{-1}$. In the visible region, the absorption peaks were strongly modulated by biaxial strain. Under compressive strain, the main absorption peaks were located at relatively shorter wavelengths, whereas tensile strain shifted the absorption response toward longer wavelengths, extending the optical activity close to the visible--IR boundary. Based on the strain-dependent band-gap values, the corresponding band-edge wavelength shifted from about $624.0~\mathrm{nm}$ at $-4\%$ strain to about $872.5~\mathrm{nm}$ at $4\%$ strain, giving a total red shift of approximately $248.5~\mathrm{nm}$. Relative to the unstrained case, the $-4\%$ compressive strain produced a blue shift of about $53.5~\mathrm{nm}$, whereas the $4\%$ tensile strain produced a red shift of about $195.0~\mathrm{nm}$. This red shift under tensile strain was consistent with the strain-induced reduction of the band gap. The refractive index also showed a clear strain-dependent behavior, with prominent peaks in the visible region. The maximum value of $\mathrm{n}$ increased to about 3.3 and gradually shifted toward longer wavelengths as the tensile strain increased. Similarly, the extinction coefficient, $\mathrm{k}$, followed the same overall trend as the absorption coefficient, since $\upalpha$ is directly proportional to $\mathrm{k}$. The peaks in $\mathrm{k}$ became more pronounced in the visible and near-IR regions under tensile strain, indicating enhanced light attenuation at longer wavelengths. Therefore, the applied biaxial strain effectively tuned the optical response of the monolayer by shifting the absorption, refractive index, and extinction coefficient spectra from the UV/visible region toward the longer-wavelength visible and near-IR regions.

\subsection{Valleytronic Device and Strain-Tunable Optoelectronic Device}

\begin{figure}[H]
    \centering
    \includegraphics[width=\textwidth]{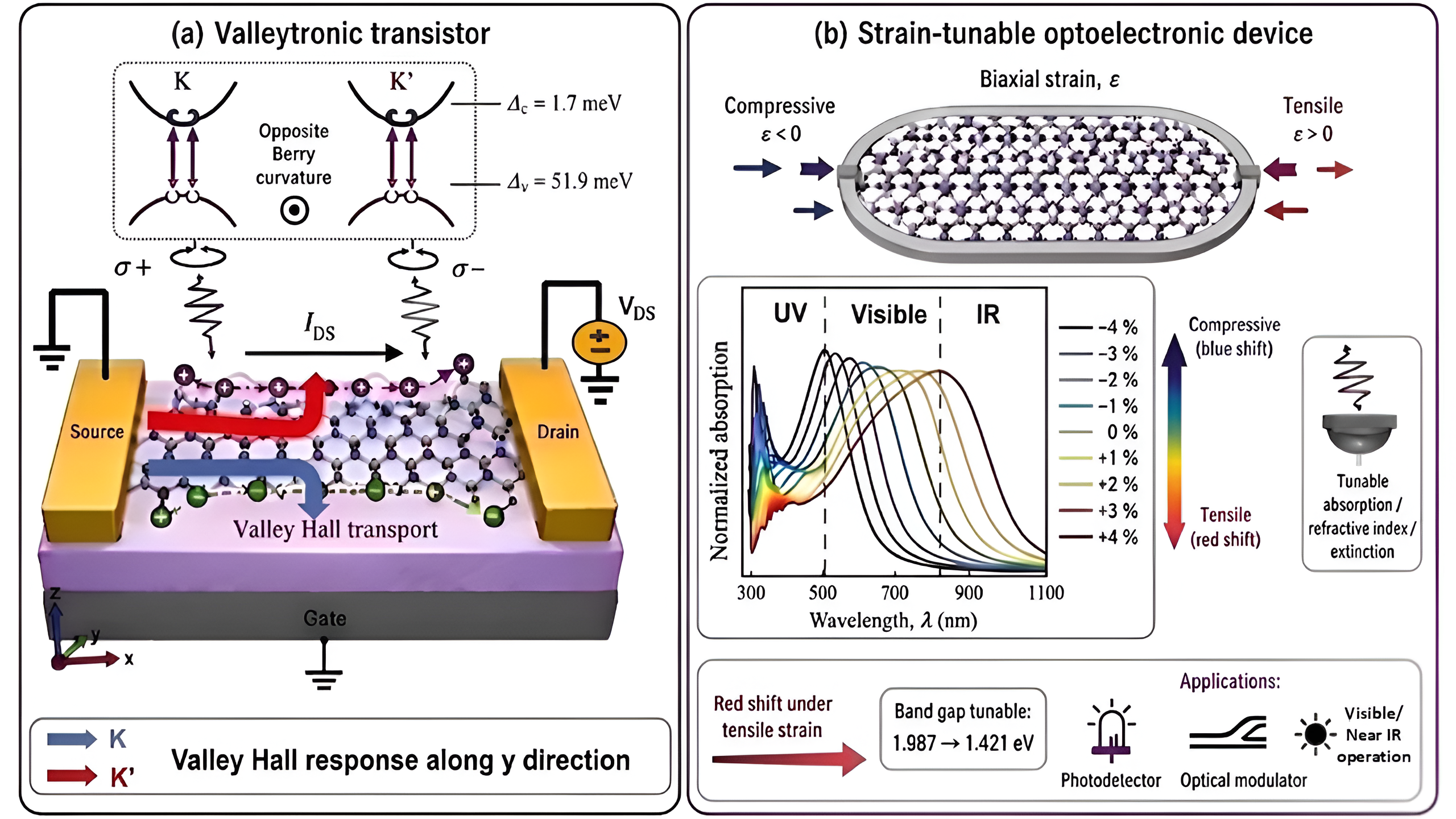}
    \caption{
    Schematic illustration of possible device applications of monolayer CrC$_2$N$_4$.
    (a) Valleytronic transistor based on valley-selective carrier transport. Circularly polarized light is expected to preferentially excite carriers at the inequivalent $\mathrm{K}$ and $\mathrm{K'}$ valleys based on valley optical selection rules, while the opposite Berry curvature at the two valleys can give rise to opposite anomalous transverse motion and possible valley accumulation at opposite sample edges.
    (b) Strain-tunable optoelectronic device based on biaxial-strain engineering. 
    Compressive and tensile strain modulate the optical response of monolayer CrC$_2$N$_4$, 
    with tensile strain producing a red shift of the absorption spectrum toward the visible and near-infrared regions.
    }
    \label{fig:application}
\end{figure}

The valley-dependent electronic and optical properties obtained in this work indicated the potential of monolayer CrC$_2$N$_4$ for valleytronic and strain-engineered optoelectronic applications, as schematically shown in \autoref{fig:application}. 
The first proposed application is a valleytronic transistor, where monolayer CrC$_2$N$_4$ acts as the conducting channel between source and drain electrodes. 
In this device concept, two energetically degenerate but momentum-distinct valleys ($\mathrm{K}$ and $\mathrm{K'}$) exist at the band edges; carriers can be selectively populated in one valley or the other, making the valley index a new information degree of freedom analogous to spin in spintronics.
The spin--orbit-coupled band structure showed valley contrasting spin polarization, with a sizable VB spin splitting of $\Delta_{\mathrm{v}} = 51.9$ meV and a smaller CB spin splitting of $\Delta_{\mathrm{c}} = 1.7$ meV. 
Under circularly polarized excitation, $\sigma^+$ and $\sigma^-$ light are expected, based on valley optical selection rules, to preferentially excite carriers in opposite valleys.
In addition, since the Berry curvature has opposite signs at the K and K$'$ valleys, the carriers acquire opposite anomalous transverse velocities under an applied in-plane electric field. 
Therefore, the proposed transistor geometry provides a possible route for converting optically generated valley polarization into an electrical signal.

The second proposed application is a strain-tunable optoelectronic device. 
The calculated results showed that monolayer CrC$_2$N$_4$ was highly sensitive to biaxial strain, making it suitable for flexible and mechanically reconfigurable optoelectronic devices. 
The band gap decreased from 1.987 eV under $-4\%$ compressive strain to 1.421 eV under $+4\%$ tensile strain. 
This strain-induced band-gap reduction shifted the corresponding band-edge wavelength from about 624.0 nm at $-4\%$ strain to about 872.5 nm at $+4\%$ strain, producing a total red shift of approximately 248.5 nm. 
Relative to the unstrained case, the $-4\%$ compressive strain produced a blue shift of about 53.5 nm, whereas the $+4\%$ tensile strain produced a red shift of about 195.0 nm. 
This shift of the band-edge transition toward longer wavelength indicates a reduction in the optical transition energy, resulting in a red shift of the absorption, refractive-index, and extinction-coefficient spectra. 
In the proposed device, mechanical strain controls the optical response of the CrC$_2$N$_4$ monolayer: compressive strain produces a blue-shifted response, whereas tensile strain shifts the response toward longer wavelengths in the visible and near-infrared regions. 
This tunable optical response makes monolayer CrC$_2$N$_4$ suitable for strain-controlled photodetectors, optical modulators, and visible/near-infrared optical components. 
Overall, the results demonstrated that monolayer CrC$_2$N$_4$ combines valley-selective optical excitation, valley contrasting Berry curvature, finite spin--valley coupling, and strain-tunable optical response within a single two-dimensional material platform. 
These features make it a promising candidate for future valleytronic transistors and mechanically tunable optoelectronic devices.

\section{Comparative Analysis}

To contextualize the valleytronic performance of monolayer CrC$_2$N$_4$, \autoref{tab:comparative_valley} presented a benchmark comparison of its band-edge splitting with those of selected representative 2D valleytronic systems. Here, \(\delta_{\mathrm{c}}\) and \(\delta_{\mathrm{v}}\) denote the magnitude of the splitting at the CB and VB edges, respectively. Depending on the system symmetry, these values correspond to either SOC-induced valley-edge spin splitting or magnetic/proximity-induced $\mathrm{K}$/$\mathrm{K'}$ valley splitting.

\begin{table}[htbp]
\centering
\caption{Comparison of valley-related band-edge splitting in CrC$_2$N$_4$ and selected representative 2D valleytronic systems. Values of \(\delta_{\mathrm{c}}\) and \(\delta_{\mathrm{v}}\) are reported in meV. For nonmagnetic systems, the values denote SOC-induced valley-edge spin splitting; for magnetic or proximity-coupled systems, they denote $\mathrm{K}$/$\mathrm{K'}$ valley splitting or valley polarization. NR denotes not reported.}
\label{tab:comparative_valley}
\footnotesize
\renewcommand{\arraystretch}{1.15}
\setlength{\tabcolsep}{5pt}
\begin{tabular*}{\columnwidth}{@{\extracolsep{\fill}}lccc@{}}
\hline
\textbf{Material} & 
\textbf{\(\delta_{\mathrm{c}}\) (meV)} & 
\textbf{\(\delta_{\mathrm{v}}\) (meV)} & 
\textbf{Reference} \\
\hline
VSi$_2$P$_4$        & 49.4         & $\sim$3          & ~~\cite{feng2021} \\
$\text{VBNS}_2$       & NR         & 48.6            & ~~\cite{long2024} \\
WS$_2$/MnO$_2$      & NR           & 43               & ~~\cite{zhou2019} \\
\textbf{CrC$_2$N$_4$}        & \textbf{1.7}          & \textbf{51.9}             & \textbf{This work} \\
CrSi$_2$N$_4$       & 26           & 130              & ~~\cite{liu2021} \\
MoSi$_2$N$_4$       & $\sim$3--17  & $\sim$130--172   & ~~\cite{li2020,ai2021,liu2023} \\
CrSi$_2$P$_4$       & 23           & 170              & ~~\cite{liu2021} \\
\hline
\end{tabular*}
\end{table}

As shown in \autoref{tab:comparative_valley}, monolayer CrC$_2$N$_4$ exhibits strongly asymmetric band-edge splitting, with a sizable VB splitting of \(\delta_{\mathrm{v}} = 51.9~\mathrm{meV}\) and a much smaller CB splitting of \(\delta_{\mathrm{c}} = 1.7~\mathrm{meV}\). The VB splitting of CrC$_2$N$_4$ exceeds those reported for VSi$_2$P$_4$ \((\delta_{\mathrm{v}}\sim 3~\mathrm{meV})\) \cite{feng2021} and the proximity-coupled WS$_2$/MnO$_2$ heterostructure \((\delta_{\mathrm{v}}=43~\mathrm{meV})\) \cite{zhou2019}. It is also slightly larger than the spontaneous valence-band valley polarization reported for ferromagnetic Janus VBNS$_2$ \((\delta_{\mathrm{v}}=48.6~\mathrm{meV})\) \cite{long2024}, indicating a clearly resolved valence-band splitting in the present nonmagnetic monolayer. However, it remains smaller than the larger splittings reported for CrSi$_2$N$_4$ \((\delta_{\mathrm{v}}=130~\mathrm{meV})\) \cite{liu2021}, MoSi$_2$N$_4$ \((\delta_{\mathrm{v}}\sim 130\text{--}172~\mathrm{meV})\) \cite{li2020,ai2021,liu2023}, and CrSi$_2$P$_4$ \((\delta_{\mathrm{v}}=170~\mathrm{meV})\) \cite{liu2021}. In contrast, the CB splitting of CrC$_2$N$_4$ is only \(\delta_{\mathrm{c}} = 1.7~\mathrm{meV}\), which is much smaller than the reported CB splittings of VSi$_2$P$_4$ \((49.4~\mathrm{meV})\) \cite{feng2021}, CrSi$_2$N$_4$ \((26~\mathrm{meV})\) \cite{liu2021}, MoSi$_2$N$_4$ \((\sim 3\text{--}17~\mathrm{meV})\) \cite{li2020,ai2021,liu2023}, and CrSi$_2$P$_4$ \((23~\mathrm{meV})\) \cite{liu2021}. Therefore, CrC$_2$N$_4$ occupies an intermediate position among the selected valleytronic systems: its VB splitting is stronger than several low- and moderate-splitting systems, whereas its CB splitting remains very small. This asymmetric behavior is consistent with the orbital character of the band edges, where the VB edge has substantial Cr-\(d_{xy}+d_{x^2-y^2}\) and N-\(p\) hybridization, while the CB edge is dominated by Cr-\(d_{z^2}\)-type states with weaker SOC-induced splitting. Although CrC$_2$N$_4$ does not exhibit the largest splitting among the compared systems, it combines finite valley-edge spin splitting, valley-contrasting Berry curvature, and strain-tunable optical response within a light-element 2D platform, making it promising for strain-engineered valleytronic and optoelectronic applications.

\section{Conclusion}

In this paper, we explored the potential of monolayer CrC$_2$N$_4$  for valleytronic and optoelectronic applications by performing electronic, valley, charge-transfer, strain-dependent, and optical properties calculations using \textit{ab initio} calculations. The optimized monolayer retained a stable septuple-layer structure and exhibited a direct band gap at the $\mathrm{K}$/$\mathrm{K'}$ valleys. The inclusion of spin--orbit coupling produced clear valley-contrasting out-of-plane spin polarization, with a sizable VB spin splitting of 51.9 meV and a much smaller CB spin splitting of 1.7 meV. Orbital-resolved analysis showed that the edge states were mainly governed by Cr-$d$ and N-$p$ hybridization, while Bader charge analysis revealed polar-covalent bonding driven by charge transfer from Cr and C atoms toward N atoms. Biaxial strain provided an effective route to tune the electronic and optical responses of CrC$_2$N$_4$. The band gap decreased from 1.987 eV under $-4\%$ compressive strain to 1.421 eV under $4\%$ tensile strain, accompanied by an indirect-to-direct band-gap transition near $-1\%$ strain. The Berry curvature remained valley contrasting at $\mathrm{K}$/$\mathrm{K'}$, and its magnitude was enhanced under tensile strain. The optical spectra further showed strong UV absorption and a strain-induced red shift of the absorption, refractive-index, and extinction-coefficient peaks toward the visible--near-infrared region. These results demonstrated that monolayer CrC$_2$N$_4$ combined valley-centered semiconducting behavior, robust spin splitting, strain-tunable Berry curvature, and tunable optical activity, making it a promising 2D candidate for strain-engineered valleytronic and optoelectronic device applications.

\begin{acknowledgement}
M.S. acknowledges financial support from the Bangladesh University of Engineering and Technology (BUET) through its Postgraduate Fellowship Program. The authors thank the Department of Electrical and Electronic Engineering, Bangladesh University of Engineering and Technology (BUET), for providing technical support and computational facilities for carrying out the first-principles calculations.
\end{acknowledgement}

\section*{Data and Code Availability Statement}
The data that support the findings of this study are available from the corresponding author
upon reasonable request.

\section*{Appendix A}
\subsection{Computational Convergence Tests}

\begin{figure}[H]
\centering
\includegraphics[width=0.90\textwidth]{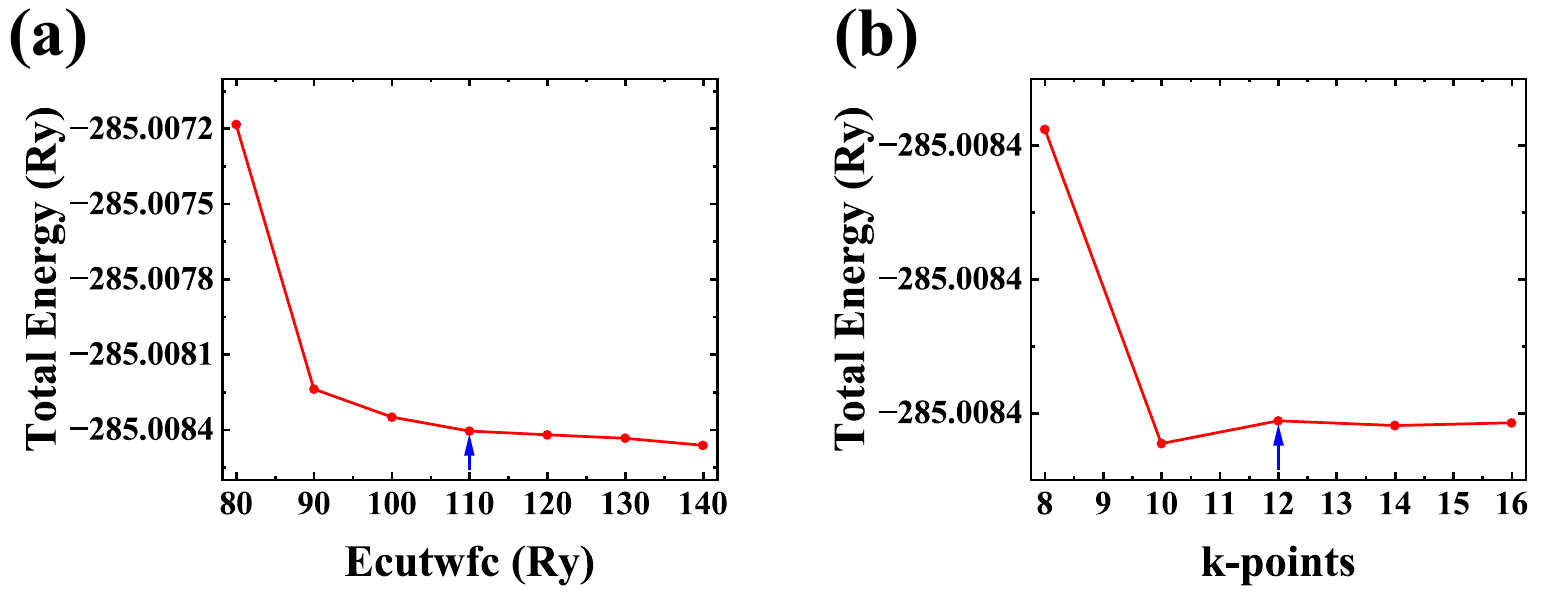}
\caption{
(a) Total energy as a function of plane-wave kinetic-energy cutoff, $E_{\mathrm{cut}}^{\mathrm{wfc}}$.  (b) Total energy as a function of the in-plane $k$-point mesh. The arrows indicate the converged values used in the calculations: $E_{\mathrm{cut}}^{\mathrm{wfc}} = 110$ Ry and a $(12 \times 12 \times 1)$ $k$ point mesh.}
\label{fig:s1}
\end{figure}

The convergence tests were performed to ensure that the calculated total energy was insensitive to the numerical parameters used in the first-principles calculations. \autoref{fig:s1} confirmed that the plane-wave cutoff energy and $k$-point mesh were converged at the selected values of  110 Ry and a $(12 \times 12 \times 1)$ grid, respectively. 

\subsection{Spin-Polarized Electronic Structure without SOC}

\begin{figure}[H]
\centering
\includegraphics[width=0.80\textwidth]{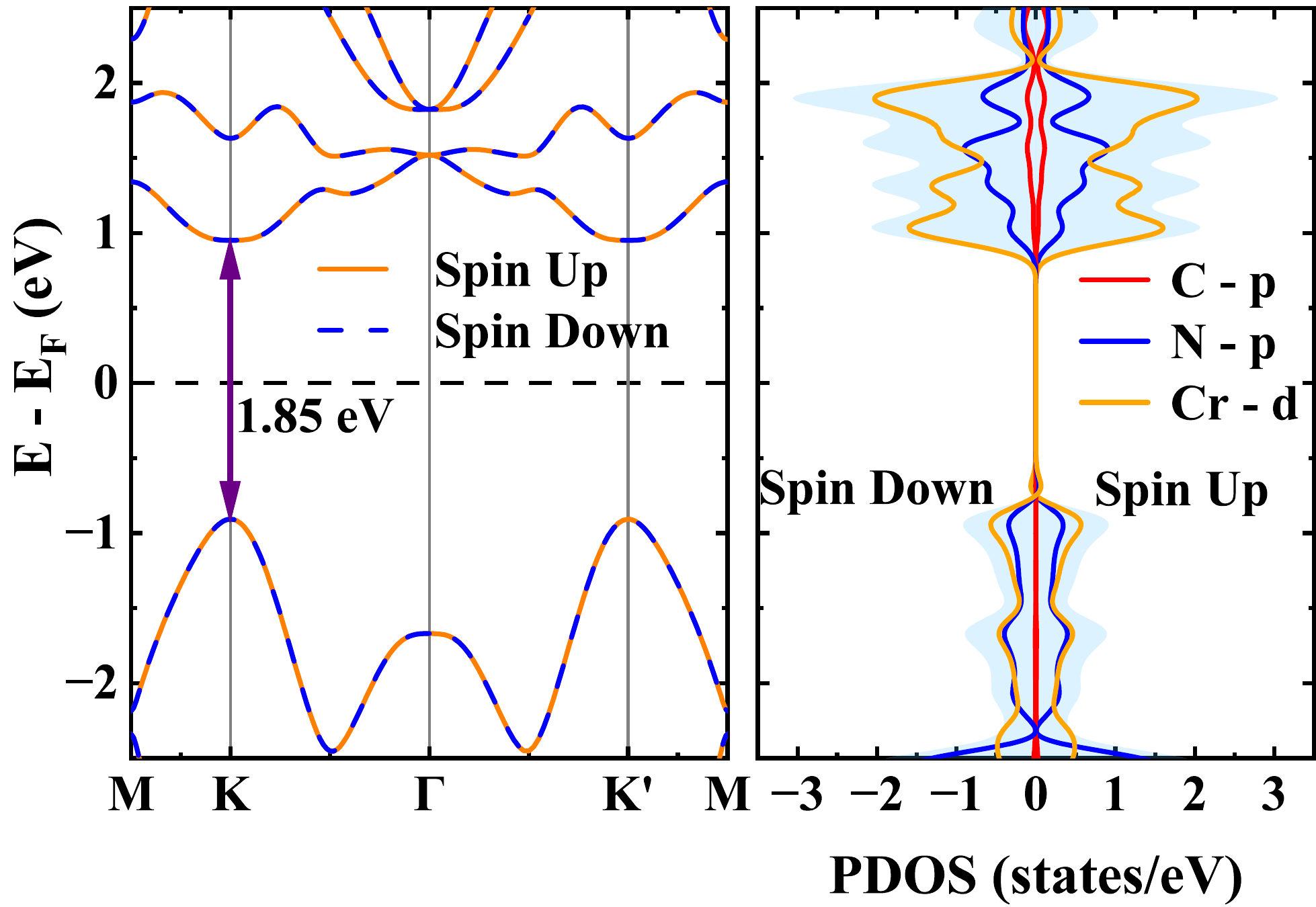}
\caption{ 
(a) Spin-resolved band structure along the M--K--$\Gamma$--K$'$--M path. The orange solid and blue dashed curves represent spin-up and spin-down channels, respectively.  (b) Spin-resolved projected density of states (PDOS), showing the orbital contributions from C-$p$, N-$p$, and Cr-$d$ states. The shaded region represents the total density of states, and the Fermi level was set to 0 eV.}
\label{fig:s2}
\end{figure}

The spin-polarized band structure and PDOS without SOC calculations were performed to verify the magnetic ground state of monolayer CrC$_2$N$_4$. As shown in \autoref{fig:s2}, the spin-up and spin-down electronic bands fully overlapped, and the corresponding PDOS showed no spin imbalance. This result confirmed the nonmagnetic character of the optimized structure before SOC was included.

\subsection{SOC Orbital-Projected Band Structures}

\begin{figure}[H]
\centering
\includegraphics[width=0.90\textwidth]{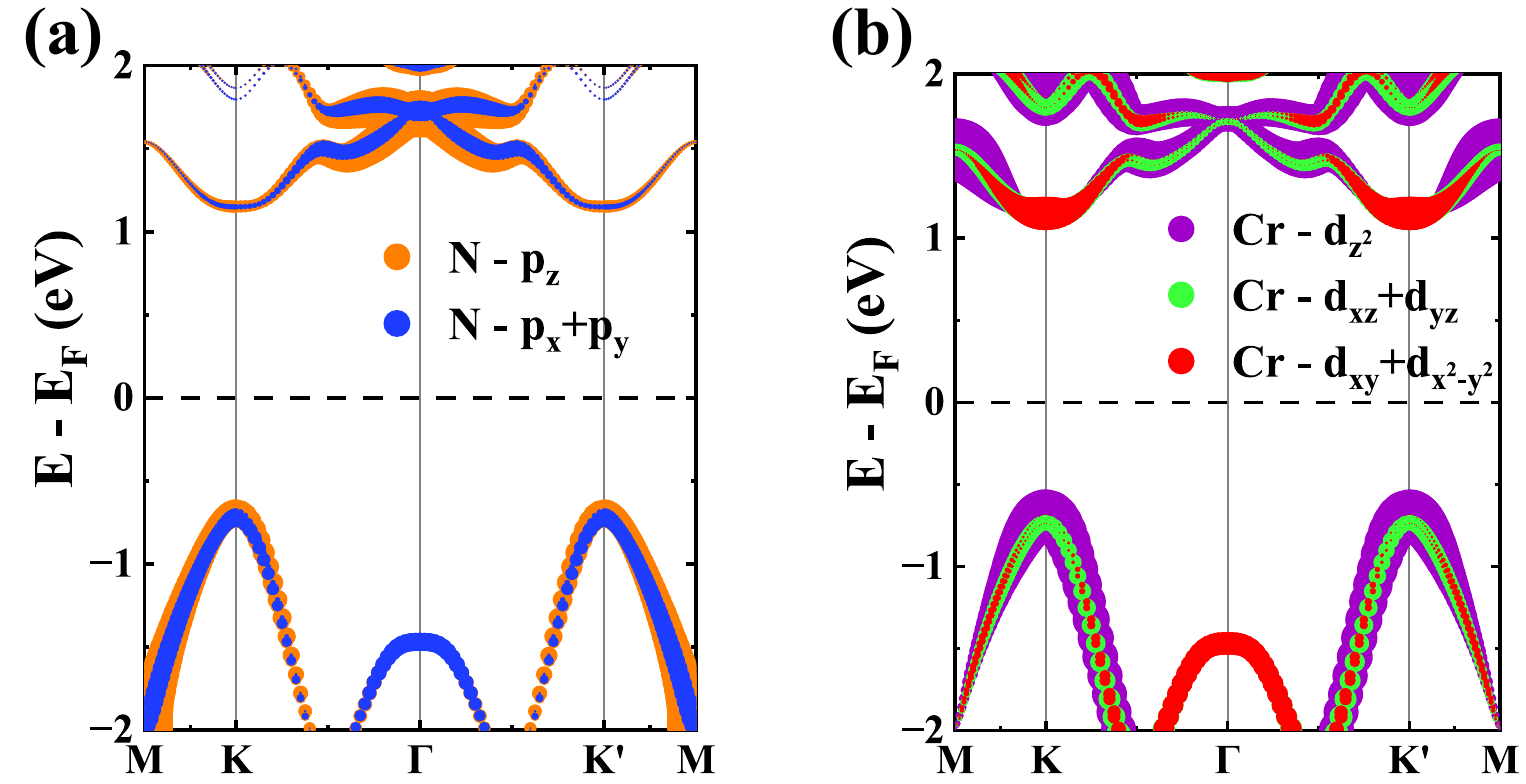}
\caption{
(a) Contributions from N-$p_z$ and N-$p_x+p_y$ orbitals. 
(b) Contributions from Cr-$d_{z^2}$, Cr-$d_{xz}+d_{yz}$, and Cr-$d_{xy}+d_{x^2-y^2}$ orbitals in monolayer CrC$_2$N$_4$. 
The symbol size represents the relative orbital weight, and the Fermi level was set to 0 eV.}
\label{fig:s3}
\end{figure}

The SOC included orbital-projected structures helps to identify the orbital origin of the valley-edge states. \autoref{fig:s3} showed that the VB edge was mainly contributed by in-plane Cr-$d_{xy}+d_{x^2-y^2}$ states with N-$p$ hybridization, whereas the CB edge was dominated by Cr-$d_{z^2}$ and Cr-$d_{xz}+d_{yz}$ components.

\section{Optimized Structural Parameters and Atomic Coordinates}

The optimized lattice parameters and fractional atomic coordinates of monolayer CrC$_2$N$_4$ are summarized in \autoref{tab:s1}. 

\begin{table}[H]
\centering
\caption{Optimized lattice parameters and fractional atomic coordinates of monolayer CrC$_2$N$_4$.}
\label{tab:s1}
\footnotesize
\renewcommand{\arraystretch}{1.15}
\setlength{\tabcolsep}{5pt}
\begin{tabular}{lcccc}
\hline
\textbf{Atom} & \textbf{Label} & \textbf{$x$} & \textbf{$y$} & \textbf{$z$} \\
\hline
Cr & Cr1 & 0.371661 & 0.697421 & 0.499987 \\
C  & C1  & 0.038316 & 0.030752 & 0.406445 \\
C  & C2  & 0.038336 & 0.030757 & 0.593531 \\
N  & N1  & 0.704981 & 0.364086 & 0.388579 \\
N  & N2  & 0.705004 & 0.364091 & 0.611397 \\
N  & N3  & 0.038324 & 0.030754 & 0.457222 \\
N  & N4  & 0.038331 & 0.030755 & 0.542753 \\
\hline
\end{tabular}

\vspace{4pt}
\footnotesize
Lattice parameters: $a=b=2.512635$~\AA, $c=28.291823$~\AA, $\alpha=\beta=90^\circ$, and $\gamma=120^\circ$.
\end{table}

\bibliography{ref}

@article{bhimanapati2015,
  title = {Recent Advances in Two-Dimensional Materials beyond Graphene},
  volume = {9},
  ISSN = {1936-086X},
  doi = {10.1021/acsnano.5b05556},
  number = {12},
  journal = {ACS Nano},
  publisher = {American Chemical Society (ACS)},
  author = {Bhimanapati,  Ganesh R. and Lin,  Zhong and Meunier,  Vincent and Jung,  Yeonwoong and Cha,  Judy and Das,  Saptarshi and Xiao,  Di and Son,  Youngwoo and Strano,  Michael S. and Cooper,  Valentino R. and Liang,  Liangbo and Louie,  Steven G. and Ringe,  Emilie and Zhou,  Wu and Kim,  Steve S. and Naik,  Rajesh R. and Sumpter,  Bobby G. and Terrones,  Humberto and Xia,  Fengnian and Wang,  Yeliang and Zhu,  Jun and Akinwande,  Deji and Alem,  Nasim and Schuller,  Jon A. and Schaak,  Raymond E. and Terrones,  Mauricio and Robinson,  Joshua A.},
  year = {2015},
  month = Nov,
  pages = {11509–11539},
}

@article{tan2017,
  title = {Recent Advances in Ultrathin Two-Dimensional Nanomaterials},
  volume = {117},
  ISSN = {1520-6890},
  doi = {10.1021/acs.chemrev.6b00558},
  number = {9},
  journal = {Chemical Reviews},
  publisher = {American Chemical Society (ACS)},
  author = {Tan,  Chaoliang and Cao,  Xiehong and Wu,  Xue-Jun and He,  Qiyuan and Yang,  Jian and Zhang,  Xiao and Chen,  Junze and Zhao,  Wei and Han,  Shikui and Nam,  Gwang-Hyeon and Sindoro,  Melinda and Zhang,  Hua},
  year = {2017},
  month = Mar,
  pages = {6225–6331},
}

@article{wu2019,
  title = {Intrinsic valley Hall transport in atomically thin ${\mathrm{MoS}}_{2}$},
  volume = {10},
  ISSN = {2041-1723},
  doi = {10.1038/s41467-019-08629-9},
  number = {1},
  journal = {Nature Communications},
  publisher = {Springer Science and Business Media LLC},
  author = {Wu,  Zefei and Zhou,  Benjamin T. and Cai,  Xiangbin and Cheung,  Patrick and Liu,  Gui-Bin and Huang,  Meizhen and Lin,  Jiangxiazi and Han,  Tianyi and An,  Liheng and Wang,  Yuanwei and Xu,  Shuigang and Long,  Gen and Cheng,  Chun and Law,  Kam Tuen and Zhang,  Fan and Wang,  Ning},
  year = {2019},
  month = Feb ,
}

@article{novoselov2020,
  title = {Discovery of 2D van der Waals layered ${\mathrm{MoSi}}_{2}{\mathrm{N}}_{4}$ family},
  volume = {7},
  ISSN = {2053-714X},
  doi = {10.1093/nsr/nwaa190},
  number = {12},
  journal = {National Science Review},
  publisher = {Oxford University Press (OUP)},
  author = {Novoselov,  Kostya S},
  year = {2020},
  month = Aug,
  pages = {1842–1844},
}

@article{hong2020,
  title = {Chemical vapor deposition of layered two-dimensional ${\mathrm{MoSi}}_{2}{\mathrm{N}}_{4}$ materials},
  volume = {369},
  ISSN = {1095-9203},
  doi = {10.1126/science.abb7023},
  number = {6504},
  journal = {Science},
  publisher = {American Association for the Advancement of Science (AAAS)},
  author = {Hong,  Yi-Lun and Liu,  Zhibo and Wang,  Lei and Zhou,  Tianya and Ma,  Wei and Xu,  Chuan and Feng,  Shun and Chen,  Long and Chen,  Mao-Lin and Sun,  Dong-Ming and Chen,  Xing-Qiu and Cheng,  Hui-Ming and Ren,  Wencai},
  year = {2020},
  month = Aug,
  pages = {670–674},
}

@article{guo2024,
  title = {Strain-induced valley polarization and quantum anomalous valley Hall effect in single septuple layer ${\mathrm{FeO}}_{2}{\mathrm{Si}}_{2}{\mathrm{N}}_{2}$},
  volume = {301},
  ISSN = {0921-5107},
  doi = {10.1016/j.mseb.2024.117193},
  journal = {Materials Science and Engineering: B},
  publisher = {Elsevier BV},
  author = {Guo,  Jiatian and Li,  Mingxin and Yuan,  Hongkuan and Chen,  Hong},
  year = {2024},
  month = Mar,
  pages = {117193},
}

@article{zhao2020,
  title = {Enhanced valley polarization at valence/conduction band in transition-metal-doped ${\mathrm{WTe}}_{2}$ under strain force},
  volume = {504},
  ISSN = {0169-4332},
  doi = {10.1016/j.apsusc.2019.144367},
  journal = {Applied Surface Science},
  publisher = {Elsevier BV},
  author = {Zhao,  X.W. and Li,  Y. and Liang,  R.D. and Hu,  G.C. and Yuan,  X.B. and Ren,  J.F.},
  year = {2020},
  month = Feb,
  pages = {144367},
}

@article{chen2021,
  title = {First-principles calculations to investigate stability,  electronic and optical properties of fluorinated ${\mathrm{MoSi}}_{2}{\mathrm{N}}_{4}$ monolayer},
  volume = {30},
  ISSN = {2211-3797},
  doi = {10.1016/j.rinp.2021.104864},
  journal = {Results in Physics},
  publisher = {Elsevier BV},
  author = {Chen,  Rui and Chen,  Dazhu and Zhang,  Weibin},
  year = {2021},
  month = Nov,
  pages = {104864},
}

@article{li2023,
  title = {Adsorption behavior of Janus ${\mathrm{MoSiGeN}}_{4}$ monolayer for gas-sensing application with high sensitivity and reuse},
  volume = {153},
  ISSN = {1386-9477},
  doi = {10.1016/j.physe.2023.115777},
  journal = {Physica E: Low-dimensional Systems and Nanostructures},
  publisher = {Elsevier BV},
  author = {Li,  Xueping and Li,  Ting and Wang,  Jianye and Song,  Xiaohui and Xia,  Congxin},
  year = {2023},
  month = Sept,
  pages = {115777},
}

@article{yadav2021,
  title = {Novel two-dimensional ${\mathrm{MA}}_{2}{\mathrm{N}}_{4}$ materials for photovoltaic and spintronic applications},
  volume = {12},
  ISSN = {1948-7185},
  doi = {10.1021/acs.jpclett.1c02650},
  number = {41},
  journal = {The Journal of Physical Chemistry Letters},
  publisher = {American Chemical Society (ACS)},
  author = {Yadav,  Asha and Kangsabanik,  Jiban and Singh,  Nirpendra and Alam,  Aftab},
  year = {2021},
  month = Oct,
  pages = {10120–10127},
}

@article{li2020,
  title = {Valley-dependent properties of monolayer ${\mathrm{MoSi}}_{2}{\mathrm{N}}_{4}, {\mathrm{WSi}}_{2}{\mathrm{N}}_{4}$, and ${\mathrm{MoSi}}_{2}{\mathrm{As}}_{4}$},
  volume = {102},
  ISSN = {2469-9969},
  doi = {10.1103/physrevb.102.235435},
  number = {23},
  journal = {Physical Review B},
  publisher = {American Physical Society (APS)},
  author = {Li,  Si and Wu,  Weikang and Feng,  Xiaolong and Guan,  Shan and Feng,  Wanxiang and Yao,  Yugui and Yang,  Shengyuan A.},
  year = {2020},
  month = Dec,
}

@article{yangjia2021,
  title = {Accurate electronic properties and non-linear optical response of two-dimensional ${\mathrm{MA}}_{2}{\mathrm{Z}}_{4}$},
  volume = {13},
  ISSN = {2040-3372},
  doi = {10.1039/d0nr09146d},
  number = {10},
  journal = {Nanoscale},
  publisher = {Royal Society of Chemistry (RSC)},
  author = {Yang,  Jia-Shu and Zhao,  Luneng and LI,  Shi-Qi and Liu,  Hongsheng and Wang,  Lu and Chen,  Maodu and Gao,  Junfeng and Zhao,  Jijun},
  year = {2021},
  pages = {5479–5488},
}

@article{guo2021,
  title = {Predicted septuple-atomic-layer Janus ${\mathrm{MSiGeN}}_{4}$ (M = Mo and W) monolayers with Rashba spin splitting and high electron carrier mobilities},
  volume = {9},
  ISSN = {2050-7534},
  doi = {10.1039/d0tc05649a},
  number = {7},
  journal = {Journal of Materials Chemistry C},
  publisher = {Royal Society of Chemistry (RSC)},
  author = {Guo,  San-Dong and Mu,  Wen-Qi and Zhu,  Yu-Tong and Han,  Ru-Yue and Ren,  Wen-Cai},
  year = {2021},
  pages = {2464–2473},
}

@article{zhou2021,
  title = {Structural Symmetry,  Spin–Orbit Coupling,  and Valley-Related Properties of Monolayer ${\mathrm{WSi}}_{2}{\mathrm{N}}_{4}$ Family},
  volume = {12},
  ISSN = {1948-7185},
  doi = {10.1021/acs.jpclett.1c03197},
  number = {48},
  journal = {The Journal of Physical Chemistry Letters},
  publisher = {American Chemical Society (ACS)},
  author = {Zhou,  Wenzhe and Wu,  Liang and Li,  Aolin and Zhang,  Bei and Ouyang,  Fangping},
  year = {2021},
  month = Nov,
  pages = {11622–11628},
}

@article{liu2023,
  title = {Characteristic excitonic absorption of ${\mathrm{MoSi}}_{2}{\mathrm{N}}_{4}$ and ${\mathrm{WSi}}_{2}{\mathrm{N}}_{4}$ monolayers},
  volume = {56},
  ISSN = {1361-6463},
  doi = {10.1088/1361-6463/ace11d},
  number = {40},
  journal = {Journal of Physics D: Applied Physics},
  publisher = {IOP Publishing},
  author = {Liu,  Hongling and Huang,  Baibiao and Dai,  Ying and Wei,  Wei},
  year = {2023},
  month = July,
  pages = {405103},
}

@article{zollner2019,
  title = {Strain-tunable orbital,  spin-orbit,  and optical properties of monolayer transition-metal dichalcogenides},
  volume = {100},
  ISSN = {2469-9969},
  doi = {10.1103/physrevb.100.195126},
  number = {19},
  journal = {Physical Review B},
  publisher = {American Physical Society (APS)},
  author = {Zollner,  Klaus and Junior,  Paulo E. Faria and Fabian,  Jaroslav},
  year = {2019},
  month = Nov,
}

@article{mortazavi2021,
  title = {Outstandingly high thermal conductivity,  elastic modulus,  carrier mobility and piezoelectricity in two-dimensional semiconducting ${\mathrm{CrC}}_{2}{\mathrm{N}}_{4}$: a first-principles study},
  volume = {22},
  ISSN = {2468-6069},
  doi = {10.1016/j.mtener.2021.100839},
  journal = {Materials Today Energy},
  publisher = {Elsevier BV},
  author = {Mortazavi,  Bohayra and Shojaei,  Fazel and Javvaji,  Brahmanandam and Rabczuk,  Timon and Zhuang,  Xiaoying},
  year = {2021},
  month = Dec,
  pages = {100839},
}

@article{liu2021,
  title = {Valley-Contrasting Physics in Single-Layer ${\mathrm{CrSi}}_{2}{\mathrm{N}}_{4}$ and ${\mathrm{CrSi}}_{2}{\mathrm{P}}_{4}$},
  volume = {12},
  ISSN = {1948-7185},
  doi = {10.1021/acs.jpclett.1c02069},
  number = {34},
  journal = {The Journal of Physical Chemistry Letters},
  publisher = {American Chemical Society (ACS)},
  author = {Liu,  Yibo and Zhang,  Ting and Dou,  Kaiying and Du,  Wenhui and Peng,  Rui and Dai,  Ying and Huang,  Baibiao and Ma,  Yandong},
  year = {2021},
  month = Aug,
  pages = {8341–8346},
}

@article{dou2020,
  title = {Promising valleytronic materials with strong spin-valley coupling in two-dimensional ${\mathrm{MN}}_{2}{\mathrm{X}}_{2}$ (M = Mo,  W; X = F,  H)},
  volume = {117},
  ISSN = {1077-3118},
  doi = {10.1063/5.0026033},
  number = {17},
  journal = {Applied Physics Letters},
  publisher = {AIP Publishing},
  author = {Dou,  Kaiying and Ma,  Yandong and Peng,  Rui and Du,  Wenhui and Huang,  Baibiao and Dai,  Ying},
  year = {2020},
  month = Oct, 
}

@article{cui2021,
  title = {Spin-valley coupling in a two-dimensional ${\mathrm{V}}{\mathrm{Si}}_{2}{\mathrm{N}}_{4}$ monolayer},
  volume = {103},
  ISSN = {2469-9969},
  doi = {10.1103/physrevb.103.085421},
  number = {8},
  journal = {Physical Review B},
  publisher = {American Physical Society (APS)},
  author = {Cui,  Qirui and Zhu,  Yingmei and Liang,  Jinghua and Cui,  Ping and Yang,  Hongxin},
  year = {2021},
  month = Feb,
}

@article{sheoran2023,
  title = {Manipulation of Valley and Spin Properties in Two-Dimensional Janus ${\mathrm{WSiGeZ}}_{4}$ (Z = N,  P,  As) through Symmetry Control},
  volume = {127},
  ISSN = {1932-7455},
  doi = {10.1021/acs.jpcc.3c02819},
  number = {23},
  journal = {The Journal of Physical Chemistry C},
  publisher = {American Chemical Society (ACS)},
  author = {Sheoran,  Sajjan and Phutela,  Ankita and Moulik,  Ruman and Bhattacharya,  Saswata},
  year = {2023},
  month = June,
  pages = {11396–11406},
}

@article{tho2023,
  title = {${\mathrm{MA}}_{2}{\mathrm{Z}}_{4}$ family heterostructures: Promises and prospects},
  volume = {10},
  ISSN = {1931-9401},
  doi = {10.1063/5.0156988},
  number = {4},
  journal = {Applied Physics Reviews},
  publisher = {AIP Publishing},
  author = {Tho,  Che Chen and Guo,  San-Dong and Liang,  Shi-Jun and Ong,  Wee Liat and Lau,  Chit Siong and Cao,  Liemao and Wang,  Guangzhao and Ang,  Yee Sin},
  year = {2023},
  month = Nov,
}

@article{ai2021,
  title = {Theoretical evidence of the spin–valley coupling and valley polarization in two-dimensional ${\mathrm{MoSi}}_{2}{\mathrm{X}}_{4}$ (X = N,  P,  and As)},
  volume = {23},
  ISSN = {1463-9084},
  doi = {10.1039/d0cp05926a},
  number = {4},
  journal = {Physical Chemistry Chemical Physics},
  publisher = {Royal Society of Chemistry (RSC)},
  author = {Ai,  Haoqiang and Liu,  Di and Geng,  Jiazhong and Wang,  Shuangpeng and Lo,  Kin Ho and Pan,  Hui},
  year = {2021},
  pages = {3144–3151},
}

@article{yao2021,
  title = {Novel Two-Dimensional Layered ${\mathrm{MoSi}}_{2}{\mathrm{Z}}_{4}$ (Z = P,  As): New Promising Optoelectronic Materials},
  volume = {11},
  ISSN = {2079-4991},
  doi = {10.3390/nano11030559},
  number = {3},
  journal = {Nanomaterials},
  publisher = {MDPI AG},
  author = {Yao,  Hui and Zhang,  Chao and Wang,  Qiang and Li,  Jianwei and Yu,  Yunjin and Xu,  Fuming and Wang,  Bin and Wei,  Yadong},
  year = {2021},
  month = Feb,
  pages = {559},
}

@article{xu2023,
  title = {Type-II ${\mathrm{MoSi}}_{2}{\mathrm{N}}_{4}$/${\mathrm{MoS}}_{2}$ van der Waals Heterostructure with Excellent Optoelectronic Performance and Tunable Electronic Properties},
  volume = {127},
  ISSN = {1932-7455},
  doi = {10.1021/acs.jpcc.3c00773},
  number = {16},
  journal = {The Journal of Physical Chemistry C},
  publisher = {American Chemical Society (ACS)},
  author = {Xu,  Xuhui and Yang,  Lei and Gao,  Quan and Jiang,  Xinxin and Li,  Dongmei and Cui,  Bin and Liu,  Desheng},
  year = {2023},
  month = Apr,
  pages = {7878–7886},
}

@article{wanglei2021,
  title = {Intercalated architecture of ${\mathrm{MA}}_{2}{\mathrm{Z}}_{4}$ family layered van der Waals materials with emerging topological,  magnetic and superconducting properties},
  volume = {12},
  ISSN = {2041-1723},
  doi = {10.1038/s41467-021-22324-8},
  number = {1},
  journal = {Nature Communications},
  publisher = {Springer Science and Business Media LLC},
  author = {Wang,  Lei and Shi,  Yongpeng and Liu,  Mingfeng and Zhang,  Ao and Hong,  Yi-Lun and Li,  Ronghan and Gao,  Qiang and Chen,  Mingxing and Ren,  Wencai and Cheng,  Hui-Ming and Li,  Yiyi and Chen,  Xing-Qiu},
  year = {2021},
  month = Apr,
}

@article{yin2023,
  title = {Emerging Versatile Two‐Dimensional ${\mathrm{MoSi}}_{2}{\mathrm{N}}_{4}$ Family},
  volume = {33},
  ISSN = {1616-3028},
  doi = {10.1002/adfm.202214050},
  number = {26},
  journal = {Advanced Functional Materials},
  publisher = {Wiley},
  author = {Yin,  Yan and Gong,  Qihua and Yi,  Min and Guo,  Wanlin},
  year = {2023},
  month = Apr, 
}

@article{jin2024,
  title = {Two-Dimensional ${\mathrm{MoSi}}_{2}{\mathrm{N}}_{4}$ Family: Progress and Perspectives Form Theory},
  volume = {15},
  ISSN = {1948-7185},
  doi = {10.1021/acs.jpclett.4c02452},
  number = {41},
  journal = {The Journal of Physical Chemistry Letters},
  publisher = {American Chemical Society (ACS)},
  author = {Jin,  Wenyuan and Zuo,  Jingning and Pang,  Jiafei and Yang,  Jinni and Yu,  Xin and Zhong,  Hongxia and Kuang,  Xiaoyu and Lu,  Cheng},
  year = {2024},
  month = Oct,
  pages = {10284–10294},
}

@article{latychevskaia2024,
  title = {A new family of septuple-layer 2D materials of ${\mathrm{MoSi}}_{2}{\mathrm{N}}_{4}$-like crystals},
  volume = {6},
  ISSN = {2522-5820},
  doi = {10.1038/s42254-024-00728-x},
  number = {7},
  journal = {Nature Reviews Physics},
  publisher = {Springer Science and Business Media LLC},
  author = {Latychevskaia,  T. and Bandurin,  D. A. and Novoselov,  K. S.},
  year = {2024},
  month = June,
  pages = {426–438},
}

@article{shu2023,
  title = {Efficient Ohmic Contact in Monolayer ${\mathrm{CrX}}_{2}{\mathrm{N}}_{4}$ (X = C,  Si) Based Field‐Effect Transistors},
  volume = {9},
  ISSN = {2199-160X},
  doi = {10.1002/aelm.202201056},
  number = {3},
  journal = {Advanced Electronic Materials},
  publisher = {Wiley},
  author = {Shu,  Yu and Liu,  Yongqian and Cui,  Zhou and Xiong,  Rui and Zhang,  Yinggan and Xu,  Chao and Zheng,  Jingying and Wen,  Cuilian and Wu,  Bo and Sa,  Baisheng},
  year = {2023},
  month = Jan,
}

@article{Giannozzi2009,
  title = {QUANTUM ESPRESSO: a modular and open-source software project for quantum simulations of materials},
  volume = {21},
  ISSN = {1361-648X},
  doi = {10.1088/0953-8984/21/39/395502},
  number = {39},
  journal = {Journal of Physics: Condensed Matter},
  publisher = {IOP Publishing},
  author = {Giannozzi,  Paolo and Baroni,  Stefano and Bonini,  Nicola and Calandra,  Matteo and Car,  Roberto and Cavazzoni,  Carlo and Ceresoli,  Davide and Chiarotti,  Guido L and Cococcioni,  Matteo and Dabo,  Ismaila and Dal Corso,  Andrea and de Gironcoli,  Stefano and Fabris,  Stefano and Fratesi,  Guido and Gebauer,  Ralph and Gerstmann,  Uwe and Gougoussis,  Christos and Kokalj,  Anton and Lazzeri,  Michele and Martin-Samos,  Layla and Marzari,  Nicola and Mauri,  Francesco and Mazzarello,  Riccardo and Paolini,  Stefano and Pasquarello,  Alfredo and Paulatto,  Lorenzo and Sbraccia,  Carlo and Scandolo,  Sandro and Sclauzero,  Gabriele and Seitsonen,  Ari P and Smogunov,  Alexander and Umari,  Paolo and Wentzcovitch,  Renata M},
  year = {2009},
  month = Sept,
  pages = {395502},
}

@article{vanSetten2018,
  title = {The PseudoDojo: Training and grading a 85 element optimized norm-conserving pseudopotential table},
  volume = {226},
  ISSN = {0010-4655},
  doi = {10.1016/j.cpc.2018.01.012},
  journal = {Computer Physics Communications},
  publisher = {Elsevier BV},
  author = {van Setten,  M.J. and Giantomassi,  M. and Bousquet,  E. and Verstraete,  M.J. and Hamann,  D.R. and Gonze,  X. and Rignanese,  G.-M.},
  year = {2018},
  month = May,
  pages = {39–54},
}

@article{Mostofi2014,
  title = {An updated version of wannier90: A tool for obtaining maximally-localised Wannier functions},
  volume = {185},
  ISSN = {0010-4655},
  doi = {10.1016/j.cpc.2014.05.003},
  number = {8},
  journal = {Computer Physics Communications},
  publisher = {Elsevier BV},
  author = {Mostofi,  Arash A. and Yates,  Jonathan R. and Pizzi,  Giovanni and Lee,  Young-Su and Souza,  Ivo and Vanderbilt,  David and Marzari,  Nicola},
  year = {2014},
  month = Aug,
  pages = {2309–2310},
}

@article{Tsirkin2021,
  title = {High performance Wannier interpolation of Berry curvature and related quantities with WannierBerri code},
  volume = {7},
  ISSN = {2057-3960},
  doi = {10.1038/s41524-021-00498-5},
  number = {1},
  journal = {npj Computational Materials},
  publisher = {Springer Science and Business Media LLC},
  author = {Tsirkin,  Stepan S.},
  year = {2021},
  month = Feb, 
}

@article{Henkelman2006,
  title = {A fast and robust algorithm for Bader decomposition of charge density},
  volume = {36},
  ISSN = {0927-0256},
  doi = {10.1016/j.commatsci.2005.04.010},
  number = {3},
  journal = {Computational Materials Science},
  publisher = {Elsevier BV},
  author = {Henkelman,  Graeme and Arnaldsson,  Andri and Jónsson,  Hannes},
  year = {2006},
  month = June,
  pages = {354–360},
}

@article{sangalli2019many,
  title = {Many-body perturbation theory calculations using the yambo code},
  volume = {31},
  ISSN = {1361-648X},
  doi = {10.1088/1361-648x/ab15d0},
  number = {32},
  journal = {Journal of Physics: Condensed Matter},
  publisher = {IOP Publishing},
  author = {Sangalli,  D and Ferretti,  A and Miranda,  H and Attaccalite,  C and Marri,  I and Cannuccia,  E and Melo,  P and Marsili,  M and Paleari,  F and Marrazzo,  A and Prandini,  G and Bonfà,  P and Atambo,  M O and Affinito,  F and Palummo,  M and Molina-Sánchez,  A and Hogan,  C and Gr\"{u}ning,  M and Varsano,  D and Marini,  A},
  year = {2019},
  month = May,
  pages = {325902},
}

@article{marini2009yambo,
  title = {yambo: An ab initio tool for excited state calculations},
  volume = {180},
  ISSN = {0010-4655},
  doi = {10.1016/j.cpc.2009.02.003},
  number = {8},
  journal = {Computer Physics Communications},
  publisher = {Elsevier BV},
  author = {Marini,  Andrea and Hogan,  Conor and Gr\"{u}ning,  Myrta and Varsano,  Daniele},
  year = {2009},
  month = Aug,
  pages = {1392–1403},
}

@article{wang2025valley,
  title = {Valley-dependent Berry phase effects and related valleytronic applications in two-dimensional materials},
  volume = {39},
  ISSN = {1793-6640},
  doi = {10.1142/s0217984925300029},
  number = {35},
  journal = {Modern Physics Letters B},
  publisher = {World Scientific Pub Co Pte Ltd},
  author = {Wang,  Sake},
  year = {2025},
  month = July, 
}

@article{wang2020strain,
  title = {Strain effect on circularly polarized electroluminescence in transition metal dichalcogenides},
  volume = {2},
  ISSN = {2643-1564},
  doi = {10.1103/physrevresearch.2.033340},
  number = {3},
  journal = {Physical Review Research},
  publisher = {American Physical Society (APS)},
  author = {Wang,  Sake and Ukhtary,  M. Shoufie and Saito,  Riichiro},
  year = {2020},
  month = Aug,
}

@book{wang2025,
  title = {Two-dimensional Valleytronic Materials: From principles to device applications},
  ISBN = {9780750355605},
  doi = {10.1088/978-0-7503-5562-9},
  publisher = {IOP Publishing},
  author = {Wang,  Sake and Tian,  Hongyu},
  year = {2025},
  month = May,
}

@article{chowdhury2026,
  title = {Strain-tunable spin filtering and valley splitting coexisting with the anomalous Hall effect in the 2D half-metallic ${\mathrm{VSe}}_{2}/{\mathrm{VN}}$ heterostructure: toward a unified spintronic--valleytronic platform},
  volume = {14},
  ISSN = {2050-7534},
  doi = {10.1039/d5tc04495b},
  number = {18},
  journal = {Journal of Materials Chemistry C},
  publisher = {Royal Society of Chemistry (RSC)},
  author = {Chowdhury,  Vivek and Zubair,  Ahmed},
  year = {2026},
  pages = {7794–7809},
}

@article{islam2026,
  title = {Tunable valley polarization and anomalous hall effect in ferrovalley ${\mathrm{NbX}}_{2}$ and ${\mathrm{TaX}}_{2}$ (X= S, Se, Te): A first-principles study},
  volume = {719},
  ISSN = {0169-4332},
  doi = {10.1016/j.apsusc.2025.165094},
  journal = {Applied Surface Science},
  publisher = {Elsevier BV},
  author = {Islam,  Samiul and Mominuzzaman,  Sharif Mohammad and Zubair,  Ahmed},
  year = {2026},
  month = Feb,
  pages = {165094},
}

@article{feng2021,
  title =  {Valley-related multiple Hall effect in monolayer ${\mathrm{V}}{\mathrm{Si}}_{2}{\mathrm{P}}_{4}$},
  ISSN = {2469-9969},
  doi = {10.1103/physrevb.104.075421},
  number = {7},
  journal = {Physical Review B},
  publisher = {American Physical Society (APS)},
  author = {Feng,  Xiangyu and Xu,  Xilong and He,  Zhonglin and Peng,  Rui and Dai,  Ying and Huang,  Baibiao and Ma,  Yandong},
  year = {2021},
  month = Aug, 
}

@article{zhou2019,
  title = {Tunable valley splitting and an anomalous valley Hall effect in hole-doped ${\mathrm{WS}}_{2}$ by proximity coupling with a ferromagnetic ${\mathrm{MnO}}_{2}$ monolayer},
  volume = {11},
  ISSN = {2040-3372},
  doi = {10.1039/c9nr03315g},
  number = {28},
  journal = {Nanoscale},
  publisher = {Royal Society of Chemistry (RSC)},
  author = {Zhou,  Baozeng and Li,  Zheng and Wang,  Jiaming and Niu,  Xuechen and Luan,  Chongbiao},
  year = {2019},
  pages = {13567–13575},
}

@article{long2024,
  title = {Characterization of ferromagnetic semiconductors and valley polarization in janus ${\mathrm{VBXS}}{_2}$ (X = N,  P) monolayers},
  volume = {99},
  ISSN = {1402-4896},
  doi = {10.1088/1402-4896/ad6ec1},
  number = {9},
  journal = {Physica Scripta},
  publisher = {IOP Publishing},
  author = {Long,  Mei and Miao,  Feng and Xu,  Min and Feng,  Shi-quan and Yang,  Yang},
  year = {2024},
  month = Aug,
  pages = {095979},
}

@article{ghosh2025,
  title = {A complementary two-dimensional material-based one instruction set computer},
  volume = {642},
  ISSN = {1476-4687},
  doi = {10.1038/s41586-025-08963-7},
  number = {8067},
  journal = {Nature},
  publisher = {Springer Science and Business Media LLC},
  author = {Ghosh,  Subir and Zheng,  Yikai and Rafiq,  Musaib and Ravichandran,  Harikrishnan and Sun,  Yongwen and Chen,  Chen and Goswami,  Mrinmoy and Sakib,  Najam U and Sadaf,  Muhtasim Ul Karim and Pannone,  Andrew and Ray,  Samriddha and Redwing,  Joan M. and Yang,  Yang and Sahay,  Shubham and Das,  Saptarshi},
  year = {2025},
  month = June,
  pages = {327–335}
}

@article{ahn2020,
  title = {2D materials for spintronic devices},
  volume = {4},
  ISSN = {2397-7132},
  doi = {10.1038/s41699-020-0152-0},
  number = {1},
  journal = {npj 2D Materials and Applications},
  publisher = {Springer Science and Business Media LLC},
  author = {Ahn,  Ethan C.},
  year = {2020},
  month = June, 
}

@article{peng2020,
  title = {Strain engineering of 2D semiconductors and graphene: from strain fields to band-structure tuning and photonic applications},
  volume = {9},
  ISSN = {2047-7538},
  doi = {10.1038/s41377-020-00421-5},
  number = {1},
  journal = {Light: Science \& Applications},
  publisher = {Springer Science and Business Media LLC},
  author = {Peng,  Zhiwei and Chen,  Xiaolin and Fan,  Yulong and Srolovitz,  David J. and Lei,  Dangyuan},
  year = {2020},
  month = Nov, 
}

@article{du2020,
  title = {Strain Engineering in 2D Material‐Based Flexible Optoelectronics},
  volume = {5},
  ISSN = {2366-9608},
  doi = {10.1002/smtd.202000919},
  number = {1},
  journal = {Small Methods},
  publisher = {Wiley},
  author = {Du,  Junli and Yu,  Huihui and Liu,  Baishan and Hong,  Mengyu and Liao,  Qingliang and Zhang,  Zheng and Zhang,  Yue},
  year = {2020},
  month = Dec, 
}

@article{yang2024,
  title = {Valleytronics Meets Straintronics: Valley Fine Structure Engineering of 2D Transition Metal Dichalcogenides},
  volume = {12},
  ISSN = {2195-1071},
  doi = {10.1002/adom.202302900},
  number = {14},
  journal = {Advanced Optical Materials},
  publisher = {Wiley},
  author = {Yang,  Shichao and Long,  Hanyan and Chen,  Wenwei and Sa,  Baisheng and Guo,  Zhiyong and Zheng,  Jingying and Pei,  Jiajie and Zhan,  Hongbing and Lu,  Yuerui},
  year = {2024},
  month = Mar, 
}

@article{schaibley2016,
  title = {Valleytronics in 2D materials},
  volume = {1},
  ISSN = {2058-8437},
  doi = {10.1038/natrevmats.2016.55},
  number = {11},
  journal = {Nature Reviews Materials},
  publisher = {Springer Science and Business Media LLC},
  author = {Schaibley,  John R. and Yu,  Hongyi and Clark,  Genevieve and Rivera,  Pasqual and Ross,  Jason S. and Seyler,  Kyle L. and Yao,  Wang and Xu,  Xiaodong},
  year = {2016},
  month = Aug, 
}

@article{xiao2012,
  title = {Coupled Spin and Valley Physics in Monolayers of ${\mathrm{MoS}}_{2}$ and Other Group-VI Dichalcogenides},
  volume = {108},
  ISSN = {1079-7114},
  doi = {10.1103/physrevlett.108.196802},
  number = {19},
  journal = {Physical Review Letters},
  publisher = {American Physical Society (APS)},
  author = {Xiao,  Di and Liu,  Gui-Bin and Feng,  Wanxiang and Xu,  Xiaodong and Yao,  Wang},
  year = {2012},
  month = May, 
}

@article{cao2012,
  title = {Valley-selective circular dichroism of monolayer molybdenum disulphide},
  volume = {3},
  ISSN = {2041-1723},
  doi = {10.1038/ncomms1882},
  number = {1},
  journal = {Nature Communications},
  publisher = {Springer Science and Business Media LLC},
  author = {Cao,  Ting and Wang,  Gang and Han,  Wenpeng and Ye,  Huiqi and Zhu,  Chuanrui and Shi,  Junren and Niu,  Qian and Tan,  Pingheng and Wang,  Enge and Liu,  Baoli and Feng,  Ji},
  year = {2012},
  month = June,
}

@article{mak2012,
  title = {Control of valley polarization in monolayer ${\mathrm{MoS}}_{2}$ by optical helicity},
  volume = {7},
  ISSN = {1748-3395},
  doi = {10.1038/nnano.2012.96},
  number = {8},
  journal = {Nature Nanotechnology},
  publisher = {Springer Science and Business Media LLC},
  author = {Mak,  Kin Fai and He,  Keliang and Shan,  Jie and Heinz,  Tony F.},
  year = {2012},
  month = June,
  pages = {494–498},
}

@article{wang2018,
  title = {Colloquium: Excitons in atomically thin transition metal dichalcogenides},
  volume = {90},
  ISSN = {1539-0756},
  doi = {10.1103/revmodphys.90.021001},
  number = {2},
  journal = {Reviews of Modern Physics},
  publisher = {American Physical Society (APS)},
  author = {Wang,  Gang and Chernikov,  Alexey and Glazov,  Mikhail M. and Heinz,  Tony F. and Marie,  Xavier and Amand,  Thierry and Urbaszek,  Bernhard},
  year = {2018},
  month = Apr, 
}

@article{luo2024,
  title = {Strong light-matter coupling in van der Waals materials},
  volume = {13},
  ISSN = {2047-7538},
  doi = {10.1038/s41377-024-01523-0},
  number = {1},
  journal = {Light: Science \& Applications},
  publisher = {Springer Science and Business Media LLC},
  author = {Luo,  Yuan and Zhao,  Jiaxin and Fieramosca,  Antonio and Guo,  Quanbing and Kang,  Haifeng and Liu,  Xiaoze and Liew,  Timothy C. H. and Sanvitto,  Daniele and An,  Zhiyuan and Ghosh,  Sanjib and Wang,  Ziyu and Xu,  Hongxing and Xiong,  Qihua},
  year = {2024},
  month = Aug,
}

@article{fujita2010,
  title = {Valley filter in strain engineered graphene},
  volume = {97},
  ISSN = {1077-3118},
  doi = {10.1063/1.3473725},
  number = {4},
  journal = {Applied Physics Letters},
  publisher = {AIP Publishing},
  author = {Fujita,  T. and Jalil,  M. B. A. and Tan,  S. G.},
  year = {2010},
  month = July, 
}

@article{wang2023,
  title = {Valley-polarized and enhanced transmission in graphene with a smooth strain profile},
  volume = {35},
  ISSN = {1361-648X},
  doi = {10.1088/1361-648x/accbf9},
  number = {30},
  journal = {Journal of Physics: Condensed Matter},
  publisher = {IOP Publishing},
  author = {Wang,  Sake and Tian,  Hongyu and Sun,  Minglei},
  year = {2023},
  month = Apr,
  pages = {304002},
}

\end{document}